\documentclass[11pt,a4paper]{article}   
\usepackage{amsmath}
\usepackage{amssymb}
\usepackage{mathrsfs}
\usepackage{mathtools}

\usepackage[numbers]{natbib}

\def\showlabelsoff{TURN OFF EQN LABELS}

\usepackage{tikz,pgfplots}
\usepackage{pstricks,pst-plot}
\usetikzlibrary{calc,patterns,decorations.pathmorphing,decorations.markings,decorations.pathmorphing,angles,quotes,arrows,arrows.meta}
\tikzset{snake it/.style={decorate, decoration=snake}}

\newlength{\figurewidth}
\newlength{\figwidth}
\newlength{\figureheight}

\newcommand{\ie}{i.e.\ }  
\newcommand{\eg}{e.g.\ } 

\newcommand{\diag}{\textrm{diag}}

\renewcommand{\exp}{{\rm exp}}

\newcommand{\bfx}{\textbf{x}}

\renewcommand{\d}{\,\mathrm{d}} 	

\renewcommand{\i}{\mathrm{i}}

\newcommand{\wrt}{\,\mathrm{d}}   
    
\newcommand{\ci}{\mathrm{i}}   
\newcommand{\e}{\mathrm{e}}

\usepackage{enumerate}

\usepackage{overpic,rotating}

\newcommand{\beq}{\begin{equation}}
\newcommand{\eeq}{\end{equation}}
\newcommand{\bea}{\begin{eqnarray}}
\newcommand{\eea}{\end{eqnarray}}
\newcommand{\bean}{\begin{eqnarray*}}
\newcommand{\eean}{\end{eqnarray*}}

\newlength{\mylinelength}
\newlength{\mydashlength}
\newlength{\mydashspace}
\newlength{\mychainlengthA}
\newlength{\mychainlengthB}
\newlength{\mychainspace}
\newlength{\mylinethickness}
\newcommand{\mathand}{\quad\textnormal{and}\quad}
\newcommand{\mathandR}{\textnormal{and}\quad}  
  
\newcommand{\mathfor}{\quad\textnormal{for}\quad}

\newcommand{\mathwhere}{\quad\textnormal{where}\quad}

\newcommand{\mathX}[1]{\quad\textnormal{#1}\quad}
  
\newcommand{\mathXL}[1]{\quad\textnormal{#1}}

\usepackage{pifont}
\ifdefined\showcomments
 \ifdefined\showlabelsoff
 \else
  \usepackage[inline]{showlabels}
  
 \fi
\fi

\ifdefined\showcomments
 \usepackage[textwidth=1.5cm,textsize=tiny]{todonotes}
 \presetkeys{todonotes}{fancyline, color=blue!30}{}
\else
 \usepackage[disable]{todonotes}
\fi 

\usepackage[normalem]{ulem}  
\usepackage{setspace}
\usepackage{xcolor}
\usepackage{marginnote}

\usepackage{soul}
\usepackage{setspace}

\usepackage{xparse}

\ifdefined\shownewold
   \newcommand{\del}[1]{{\color{red}\sout{#1}}}
\else
   \newcommand{\del}[1]{\ignorespaces}
\fi
\ifdefined\shownewold
   \newcommand{\deleqn}[1]{\\\del{\parbox{\textwidth}{#1}}}
\else
   \newcommand{\deleqn}[1]{\ignorespaces}
\fi
\ifdefined\shownewold
   \newcommand{\new}[1]{{\color[rgb]{0,0.5,0}#1}}
\else
   \newcommand{\new}[1]{#1}
\fi
\ifdefined\shownewold
   \newcommand{\old}[1]{{\color[rgb]{0.5,0.5,0.5}#1}}
\else
   \newcommand{\old}[1]{\ignorespaces}
\fi
\ifdefined\shownewold
   \newcommand{\delref}[1]{\del{\parbox{\figwidth\columnwidth}{#1}}}
\else
   \newcommand{\delref}[1]{\vspace{-12pt}}
\fi
\ifdefined\showcomments
   \newcommand{\lu}[1]{{\color[rgb]{0.7,0,0.7} \textit{#1}}}
\else
   \newcommand{\lu}[1]{\ignorespaces}
\fi

\ifdefined\shownewold
 \usepackage{graphicx}
 \usepackage[many,skins]{tcolorbox}
 \newtcolorbox{cross}{blank,breakable,parbox=false,enhanced,
  finish={\draw[red,line width=2pt] (interior.south west)--(interior.north east);
    \draw[red,line width=2pt] (interior.north west)--(interior.south east);}}
  \newcommand{\figcross}[1]{\begin{tikzpicture}
                            \node (mypic) [anchor=south west, inner sep=0pt] {#1};
                            \draw[-,red,ultra thick] (mypic.south west) -- (mypic.north east);
                            \draw[-,red,ultra thick] (mypic.north west) -- (mypic.south east);
                            \end{tikzpicture}}
  \newenvironment{figurecross}{\begin{center}}{\end{center}}  
\else
 \usepackage{environ}
 \usepackage{graphicx}
 \usepackage[many,skins]{tcolorbox}
 \newtcolorbox{cross}{blank,breakable,parbox=false,enhanced,
  finish={\draw[red,line width=5pt] (interior.south west)--(interior.north east);
    \draw[red,line width=5pt] (interior.north west)--(interior.south east);}}
  \RenewEnviron{cross}{\ignorespaces}    
 \newcommand{\figcross}[1]{\ignorespaces}  
 \newenvironment{figurecross}{\begin{figure}}{\end{figure}}
 \RenewEnviron{figurecross}{\ignorespaces} 
\fi

\usepackage{xcolor}
\definecolor{darkgreen}{rgb}{0,0.55,0}
\definecolor{midgreen}{rgb}{0,0.8,0.2}
\definecolor{magenta}{rgb}{1,0,1}
\definecolor{purple}{rgb}{0.5,0,0.5}
\definecolor{darkorange}{rgb}{1,0.55,0}
\definecolor{maroon}{rgb}{0.5,0,0}
\definecolor{olive}{rgb}{0.5,0.5,0}
\definecolor{midgrey}{rgb}{0.5,0.5,0.5}
\definecolor{lightgrey}{rgb}{0.75,0.75,0.75}
\definecolor{matlabblue}{rgb}{0,0.447,0.741}
\definecolor{matlabred}{rgb}{0.85,0.325,0.098}
\definecolor{lightblue}{rgb}{0,0.5,1}
\definecolor{darkgrey}{rgb}{0.25,0.25,0.25}
\definecolor{teal}{rgb}{0,0.5,0.5}
\definecolor{navy}{rgb}{0,0,0.5}
\definecolor{goldenrod}{rgb}{0.85,0.6,0.1}

\begin{document}


\title{Swell induced vibrations of a thickening ice shelf over a shoaling seabed}

\author{M.~H.~Meylan$^{1}$,
M.~Ilyas$^{1}$, 
B.~P.~Lamichhane$^{1}$
and
L.~G.~Bennetts$^{2}$
\\
{\footnotesize
$^{1}$School of Mathematical and Physical Sciences, The University of Newcastle, Callaghan, NSW 2308, Australia}
\\
{\footnotesize
$^{2}$School of Mathematical Sciences, University of Adelaide, Adelaide, SA 5005, Australia
}}
\date{} 

\maketitle

\begin{abstract}
A solution method is developed for a model of ice shelf vibrations in response to ocean waves, 
in which the ice shelf thickness and seabed beneath the ice shelf vary over distance, and the ice shelf/sub--ice--shelf cavity are connected to the open ocean.
The method combines a decomposition of the ice shelf motion into 
free modes of vibration, 
a finite element method for the cavity water motion, 
and a non-local operator to connect to the open ocean.
An investigation is conducted into the effects of ice shelf thickening, seabed shoaling and \del{viscoelastic damping}
\new{the grounding-line conditions}
on ice shelf vibrations, 
induced by regular incident waves in the swell regime.
Further, results are given for ice shelf vibrations in response to irregular incident waves, and ocean-to-ice-shelf transfer functions are derived.
The findings add to evidence that ice shelves experience appreciable vibrations in response to swell,
and that ice shelf thickening and seabed shoaling can have
a considerable influence on predictions of how ice shelves respond to ocean waves.
\end{abstract}

\section{Introduction}
\label{sec:intro}

Seismic measurements from ice shelves (and ice tongues) around the Antarctic coastline show \del{that} they vibrate in response to ocean surface waves
\cite{macayeal2006transoceanic,cathles2009seismic,bromirski2010transoceanic,Bro&Ste12,Broetal15,bromirski2017tsunami}, 
ranging from regularly-occurring long-period swell (in summer when the sea ice barrier surrounding the shelves is \del{weakest}\new{weak}
or absent) to episodic infragravity and tsunami waves.
There is strong evidence \del{that} the vibrations cause icequakes \cite{Bro&Ste12,chen2019ross}, major iceberg calvings \cite{macayeal2006transoceanic,brunt2011antarctic}, and even trigger catastrophic disintegration events \citep{massom_etal18}. 
Damage to ice shelves has broad implications for the environment,
as ice shelves mitigate
discharge of land-based ice into the ocean \cite{gudmundsson2013}. 
The rate of glacier outflow has been found to accelerate up to eight--fold following \del{the} disintegration of adjoining ice shelves \cite{scambos2004glacier},
making future weakening/loss of ice shelves a significant concern for rising global sea levels.

Measurements of ice shelf vibrations have been accompanied by model developments, based on the hydroelastic framework set by \citep{holdsworth1978iceberg,holdsworth1981mechanism,vinogradov1985oscillation,fox1991coupling}, in which the shelf is modelled as a (thin) elastic body floating on an inviscid fluid (the sub-shelf water cavity), and, typically, clamped at the \del{landward end}
\new{grounding line (the point at which the shelf starts to float)} and free at the shelf front (its seaward end).
\new{The free end conditions neglect the bending moment imposed on the submerged portion of the shelf front by water pressure,
	although numerical simulations indicate the pressure has only a small effect on shelf vibrations \cite{kalyanaraman_coupled_2020}.}
\new{The clamped edge conditions assume the shelf is fixed to a stiff bed at the grounding line. 
 	They were employed in the early ice-shelf vibration models proposed by \cite{holdsworth1978iceberg,holdsworth1981mechanism}, 
	consistent with contemporary models of tidal interactions with ice shelves \cite{holdsworth1977tidal}.
	Hinged edge grounding line conditions were also suggested by \cite{holdsworth1981mechanism}, but they stated the clamped condition 
	\textit{is more nearly the correct condition in nature}.}
\new{Some modern tide--shelf interaction models allow grounding line migration (movement due to tides, as supported by observations) by assuming a soft bed and free ice shelf connection \cite{sayag_elastic_2013}, 
	    whilst others neglect migration and use boundary conditions analogous to clamping \cite{rosier_interpretation_2017}.}

  One class of \new{ice shelf vibration} model treats only the \del{coupled} shelf--cavity system, assuming some \del{prescribed} \new{simple} coupling with the open ocean, which is likely to be valid only for very long incident ocean waves that propagate freely into the shelf--cavity region. 
The simplest variant of the model is a \new{one-dimensional} uniform shelf on a \new{one-dimensional} shallow-water cavity with a flat seabed\new{, which results in cavities being terminated at the grounding line by unrealistically large, vertical submarine cliffs.} \del{, both in one (horizontal) dimension.} Imposing a fluid flux condition at the seaward end of the cavity leads to an eigen\del{--}problem,  and estimates of \del{the} 
resonant frequencies and mode shapes \cite{Ser13,meylan_etal17_annals}. 
A more sophisticated version of the model uses the full linear elasticity equations for the shelf and a prescribed wave forcing. It has been implemented in one horizontal dimension (plus a vertical dimension for the shelf) for uniform geometry, using a separation of variables solution technique, and used to study the stresses induced by a handful of long waves \cite[periods $>60$\,s;][]{Sergienko2010}. It has also been implemented in two horizontal dimensions for variable thickness and seabed profiles representative of the Ross Ice Shelf, using finite element package software, and results given for simulations with long incident waves ($100$\,s and $300$\,s periods) and different incident directions \cite{sergienko2017behavior}.

\del{The}\new{A} second class of model incorporates interactions between the shelf--cavity region and the open ocean, in which an incident wave is prescribed, and into which the shelf reflects wave energy. This class of model can be used to predict the proportion of the incident wave that penetrates the shelf--cavity region, \ie it is valid in the swell regime. In a series of papers \cite{papathanasiou_higher_2015,papathanasiou2015hydroelastic,papathanasiou2019resonant}, a finite-element solution method was developed for a uniform shelf on shallow water with a varying seabed, in one spatial dimension. They showed the ice shelf resonates at periods close to those predicted by the shelf--cavity model, and that seabed shoaling leading up to the shelf 
reflects a proportion of the incident waves before they reach the shelf.  
However, although wavelengths in the shelf--cavity region are much longer than in the open ocean, 
for accurate predictions of reflection by the shelf and shelf vibrations
it is necessary to use a finite-depth model, 
in  which water motion depth variations are resolved,
when the shallow-water assumption no longer holds in the open ocean
\cite[periods below $\approx{}20$\,s for 100\,m water depth;][]{kalyanaraman2019shallow}.
A solution method for a uniform shelf on finite-depth water has been developed, using a modal decomposition for the shelf vibrations in terms of its in-vacuo modes (\ie uncoupled from the fluid and independent of the forcing frequency), and a finite element method in the cavity that accommodates varying seabeds \cite{ilyas2018time}. 
Results indicate the bed profile influences the resonant responses of the shelf to wave forcing.

Ice shelf thickness profiles are highly variable, but tend to thicken with distance away from the shelf front towards the \del{landward end}\new{grounding line}---see, e.g., Fig.~7 in \cite{sergienko2017behavior} and Fig.~1 in \cite{kalyanaraman2019shallow}, which are obtained from the BEDMAP2 dataset \cite{bedmap2}.
 Intuitively, shelf thickening is expected to localise vibrations towards the shelf front, particularly for short period waves, and alter the resonant structure, which depends on interactions at its landward end \cite{kalyanaraman2019shallow}. In this article, we outline (to our knowledge) the first solution method for a shelf of varying thickness on finite-depth water with a varying seabed, including the coupling between the shelf--cavity region and the open ocean. 
 \new{The thickening and shoaling reduce/remove the unrealistically large, vertical submarine cliff at the grounding line.}
 Similar to \cite{ilyas2018time}, the method is based on an expansion of the shelf vibrations in its in-vacuo modes, but in the more challenging setting of an inhomogeneous shelf where analytical expressions for the modes are not readily available. The method proposed is efficient in the sense that only a small number of modes are required for accurate approximations (order ten for typical situations),
and that the modes are expressed in terms of uniform-shelf modes, so that the inner-products needed to couple the shelf to the underlying water cavity only need to be calculated for the uniform modes, and can be reused for any shelf profile. 
\new{As will be shown, the modal solution method provides insights into the ice shelf resonances.}
We use the method to study shelf displacement and strain profiles, and the resonant structure over for a spectrum of periods in the swell regime (10--50\,s), 
for shelves that thicken at different gradients, over seabeds that shoal at different gradients.
\new{Further, we construct ocean-to-shelf transfer functions for displacement and strain in response to irregular incident wave fields, and compare responses for clamped and hinged shelves.}

\section{Problem formulation}
\label{sec:problem}

\begin{figure}[h!]
        \centering
        \setlength{\figurewidth}{0.85\textwidth} 
        \setlength{\figureheight}{0.5\textwidth} 
        \begin{cross}
%
\definecolor{mycolor1}{rgb}{0.00000,1.00000,1.00000}%
\begin{tikzpicture}

\begin{axis}[%
width=0.951\figurewidth,
height=\figureheight,
at={(0\figurewidth,0\figureheight)},
scale only axis,
xmin=-2000,
xmax=4250,
xtick={0,4000},
xlabel style={font=\color{white!15!black}},
xticklabels={0,$L$},
xlabel={$x$},
ymin=-210,
ymax=50,
ytick={-200,-40,0},
yticklabels={$-h_{0}$,$-d$,0},
ylabel style={font=\color{white!15!black}},
ylabel={$z$},
axis background/.style={fill=white},
axis x line*=bottom,
axis y line*=left
]

\fill[mycolor1!35, domain=0:4000, variable=\x, samples=200] (axis cs:-2000,-200) -- (axis cs:0,-200) -- plot (axis cs: {\x},{100*(\x/4000)^2 - 200}) -- (axis cs:4000,-40) -- (axis cs: 0,-40) -- (axis cs: 0,0) -- (axis cs: -2000,0) --cycle;

\fill[gray!35, draw = black, line width=2.0pt, domain=0:4000, variable=\x, samples=200] (axis cs:4000,-40) -- (axis cs:0,-40) -- (axis cs: 0,10) -- plot (axis cs: {\x},{10+30*tanh(3.14*\x/4000)}) -- (axis cs:4000,-40) --cycle;

\draw[black, line width=2.0pt, domain=0:4000, variable=\x, samples=200] (axis cs: -4000,-200) -- (axis cs:0,-200) -- plot (axis cs: {\x},{100*(\x/4000)^2 - 200}) node[below,xshift=-54pt,yshift=-42pt] {$z=-h(x)$} -- (axis cs:4000,-40) -- (axis cs: 0,-40) -- (axis cs:0,0) -- (axis cs: -2000,0);

\node[rotate=-90] at (axis cs: 4150, -40) {land};

\node at (axis cs: 1000, -5) {shelf};
\node[align=center] at (axis cs: 1000, -100) {water cavity \\ $\Omega_{\textnormal{c}}$};
\node[align=center] at (axis cs: -1000, -100) {open ocean \\ $\Omega_{\textnormal{o}}$};

\draw[latex-latex,black,thick] (axis cs:2500,-40) -- node[right] {$H(x)$} (axis cs:2500,39);

\addplot [color=matlabred, line width=1.0pt, dotted]
    table[row sep=crcr]{%
0	-40\\
0   -200\\
};

%
%








\end{axis}

\end{tikzpicture}%
        \end{cross}
        
%
\definecolor{mycolor1}{rgb}{0.00000,1.00000,1.00000}%
\begin{tikzpicture}

\begin{axis}[%
width=0.951\figurewidth,
height=\figureheight,
at={(0\figurewidth,0\figureheight)},
scale only axis,
xmin=-2000,
xmax=4250,
xtick={0,4000},
xlabel style={font=\color{white!15!black}},
xticklabels={0,$L$},
xlabel={$x$},
ymin=-210,
ymax=50,
ytick={-200,0},
yticklabels={$-h_{0}$,0},
ylabel style={font=\color{white!15!black}},
ylabel={$z$},
axis background/.style={fill=white},
axis x line*=bottom,
axis y line*=left
]


\fill[mycolor1!35, domain=0:4000, variable=\x, samples=200] (axis cs:-2000,-200) -- (axis cs:0,-200) -- plot (axis cs: {\x},{70*(\x/4000)^2 - 200}) -- (axis cs:4000,-40) -- (axis cs: 0,-40) -- (axis cs: 0,0) -- (axis cs: -2000,0) --cycle;



\fill[gray!35, domain=0:4000, variable=\x, samples=200] 
    plot (axis cs: {\x},{-105+65*tanh(3.14*(4000-\x)/4000)}) --
    (axis cs:4000,20) --
    plot (axis cs: {4000-\x},{20-10*tanh(3.14*\x/4000)}) --cycle;



\draw[black, line width=2.0pt, domain=0:4000, variable=\x, samples=200] (axis cs: -4000,-200) -- (axis cs:0,-200) -- plot (axis cs: {\x},{70*(\x/4000)^2 - 200}) node[below,xshift=-54pt,yshift=-32pt] {$z=-h(x)$} 
	-- plot (axis cs: {4000-\x},{-105+65*tanh(3.14*(\x)/4000)})
	node[below,xshift=80pt,yshift=-5pt] {$z=-d(x)$}
	-- plot (axis cs: {\x},{20-10*tanh(3.14*(4000-\x)/4000)});
	
\draw[black, line width=2.0pt] 
	(axis cs:0,0) -- (axis cs: -2000,0);
	
\draw[black, dashed, line width=2.0pt] 
	(axis cs:4000,20) -- (axis cs: 4000,-105);
	


\node at (axis cs: 1000, -15) {ice shelf};
\node[align=center] at (axis cs: 1000, -120) {water cavity \\ $\Omega_{\textnormal{c}}$};
\node[align=center] at (axis cs: -1000, -120) {open ocean \\ $\Omega_{\textnormal{o}}$};

\draw[latex-latex,black,thick] (axis cs:2500,-51.25) -- node[right] {$H(x)$} (axis cs:2500,10.38);

\draw[latex-latex,black,thick] (axis cs:-100,0) -- node[left] {$d_{0}$} (axis cs:-100,-40);

\draw[latex-latex,black,thick] (axis cs:3900,-105) -- node[left, align = center] {submarine \\ cliff} (axis cs:3900,-130);

\addplot [color=matlabred, line width=1.0pt, dotted]
    table[row sep=crcr]{%
0	-40\\
0   -200\\
};

\end{axis}

\end{tikzpicture}%
        \caption{Schematic (not to scale) of the geometry.}
        \label{fig:mode_shapes}
\end{figure}
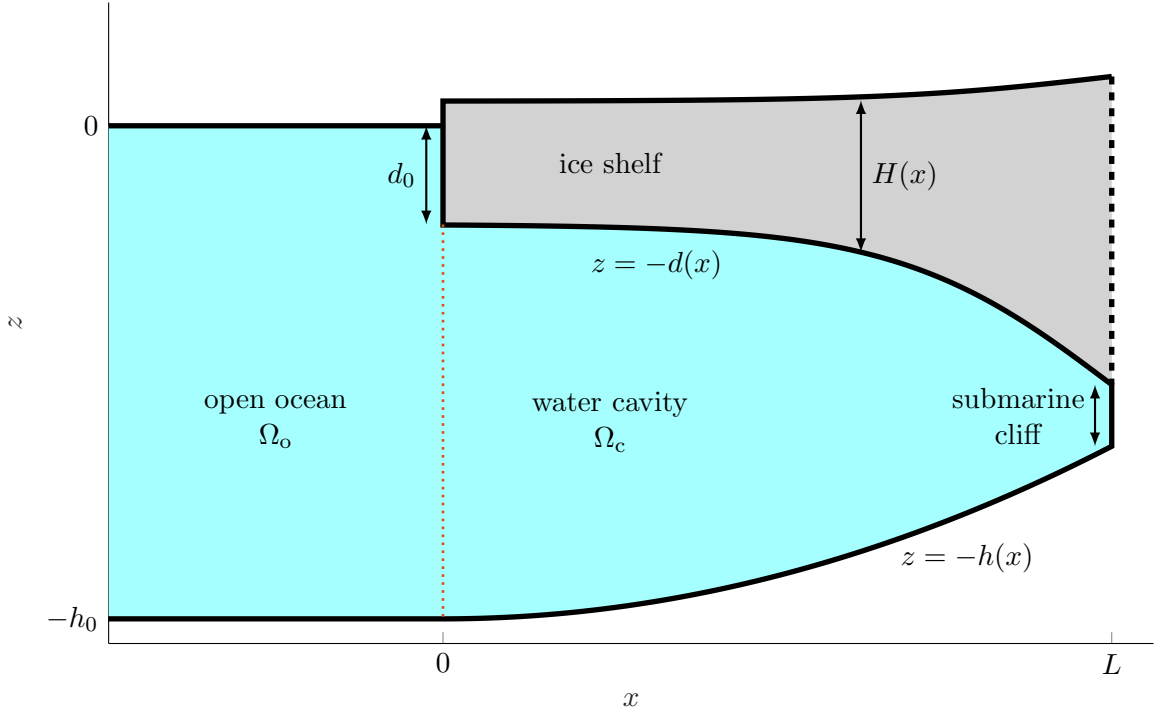

Consider a two-dimensional model with Cartesian coordinates $\mathbf{x}=(x,z)$,
where $x$ is the horizontal coordinate and $z$ is the vertical coordinate.
The open ocean occupies the region 
\begin{equation}
\mathbf{x}\in\Omega_\textnormal{o}=\{(x,z): x<0, -h_0<z<0\},
\end{equation}
where $z=0$ is the equilibrium ocean surface,
and $z=-h_0$ is the location of the flat impermeable seabed. 
The sub-shelf water cavity
occupies the region 
\begin{equation}
\mathbf{x}\in\Omega_\textnormal{c}=\{(x,z): 0<x<L, -h(x)<z<-d\},
\end{equation}
where $z=-d\new{(x)}$ is the location of the ice shelf underside \del{(assumed flat for simplicity)}, 
and $z=-h(x)$ is the location of the seabed beneath the shelf, which joins continuously to the seabed in the open ocean region, \ie $h(0)=h_{0}$,
and meets land at $x=L$.

Under the standard assumptions of linear water-wave theory (incompressible, inviscid fluid, undergoing small amplitude, irrotational motion), 
the water velocity field is defined 
as the gradient of the \del{a} velocity potential $\Phi(\mathbf{x},t)$, where $t$ denotes time.
The velocity potential satisfies Laplace's equation throughout the water, 
\begin{equation}
    \Big(\partial_{x}^{2}+\partial_{z}^{2}\Big)\,\Phi(\mathbf{x},t) =0
    \mathfor
    \mathbf{x}\in\,\Omega_{\textnormal{o}}\cup \Omega_{\textnormal{c}},
\end{equation}
where $\partial_{\bullet}\equiv{}\partial\,/\,\partial{}\bullet$,
no-normal-flow conditions on the ice shelf front, land and seabed, respectively,
\begin{align}
    & \partial_x\Phi  = 0
    \mathfor
    x=0 \mathand -d<z<0,
    \\
    & \partial_x\Phi  = 0
    \mathfor
    x=L \mathand -h(L)<z<-d,
    \\
    \mathandR &
    \left\{
    \begin{array}{ll}
    \partial_z\Phi = 0  
    &\mathfor
    x<0 \mathand z=-h_{0},
    \\[5pt]
    \partial_n\Phi = 0 
    & \mathfor
    0<x<L \mathand z = -h(x), 
    \end{array}
    \right.
\end{align}
where $n$ denotes the normal direction,
and the open ocean free-surface conditions
\begin{equation}\label{eq:freesurf}
    \partial_{t}\Phi+g\,u=0
    \mathand
    \partial_{z}\Phi=\partial_{t}u
    \mathfor
    z=0
    \mathand 
    x<0,
\end{equation}
where $u(x,t)$ is the free-surface vertical displacement,
and $g=9.81${\,m\,s$^{-2}$} is the constant of gravitational acceleration.
The two components of equation~\eqref{eq:freesurf} can be combined to leave the condition
\begin{equation}
    \partial_{t}^{2}\Phi + g\,\partial_z \Phi = 0
    \mathfor
    z=0
    \mathand 
    x<0,
\end{equation}
in which the velocity potential is the only unknown.

The ice shelf is modelled as a thin elastic plate, 
with thickness $H(x)$, 
density $\rho_{\textnormal{i}}=917$\,kg\,m$^{-3}$, 
Young's modulus $E=11$\,GPa,
and Poisson's ratio $\nu=0.3$.
\new{The (effective) Young's modulus of ice shelves is debated, 
with experimental measurements (approximately the value used)
up to an order of magnitude greater than
model best-fits to tidal flexure data
\cite{vaughan_tidal_1995}.
However, \cite{sayag_elastic_2013} show the discrepancy can be removed by using 
the soft-bed/free-shelf model rather than the stiff-bed/fixed-shelf model. 
In the present model, 
using a lower value of Young's modulus would decrease the wavelength in the shelf--cavity region, and can be interpreted as a period up-shift in the results presented.}

The vertical shelf displacement, $u(x,t)$ ($0<x<L$),
obeys the equation of motion
\begin{equation}\label{eq:rigidity}
    \partial_x^{2}\left(F(x)\,\partial_x^{2}u\right) + 
    m(x)\,
    \partial_t^{2}u=
    -\rho_{\textnormal{w}}\,\{
    g\,u
    +
    \partial_t\Phi\}
    \mathfor
    z=-d\new{(x)}
    \mathand 
    0<x<L,
\end{equation}
where $\rho_{\textnormal{w}}=1027$\,kg\,m$^{-3}$ is the water density, $m(x)=\rho_{\textnormal{i}}\,H(x)$ is the shelf mass per unit area, and
\begin{equation}
F(x)= \frac{E\,H(x)^{3}}{12\,(1-\nu^{2})}
\mathXL{is the shelf rigidity.}
\end{equation}
The shelf displacement and velocity potential
are also coupled by the kinematic condition
\begin{equation}
    \partial_{t}u=\del{\partial_{z}\Phi}
    \new{\partial_{n}\Phi}
    \mathfor
    z=-d\new{(x)}
    \mathand 
    0<x<L.
\end{equation}
Free-edge conditions are applied at the shelf front,  
\begin{subequations}
\begin{equation}\label{boundary}
    \partial_{x}^{2}u =\partial_{x}^{3}u=0
    \mathfor x=0_{+},
\end{equation}
and clamped conditions are applied at \del{landward end} \new{the grounding line}    
\begin{equation}
    \del{u  = \partial_{x}u=0}
    \new{u = 0
    \mathand 
    \partial_{x}u=0}
    \mathfor
    x=L.
\end{equation}
\end{subequations}
\new{Results for hinged conditions at the grounding line, rather than clamped conditions, \ie
\begin{equation}
	 u  = 0
	 \mathand
	 \partial_{x}(F\,\partial_{x}u)=0
    \mathfor
    x=L,
\end{equation}
are also shown, where the second condition represents vanishing of bending moment.}

The problem is mapped from the time domain to the frequency domain by expressing the unknown functions, $\Phi(x,z,t)$ and $u(x,t)$,  as
\begin{equation}
    \Phi\left(x,z,t\right) =\operatorname{Re}
    \left\{\phi(x,z:\omega)\, \e^{-\i\,\omega\, t}\right\}
    \mathand
    u\left(x,t\right) =\operatorname{Re}\left\{\eta(x:\omega) \,\e^{-\i\,\omega\,t}\right\},
\end{equation}
where $\phi$, $\eta\in\mathbb{C}$,
and $\omega\in\mathbb{R}>0$ is angular frequency.
An incident wave from the open ocean is applied via the radiation condition
\begin{equation}\label{eq:radcond}
    \phi(x,z) \sim 
    \{
    a(\omega)\,\e^{\ci\,k\,x}
    +
    b(\omega)\,\e^{-\ci\,k\,x}\}\,w(z)
    \mathX{as} x\to-\infty,
\end{equation}
where the wavenumber $k$ is the positive real root of the dispersion relation
\begin{equation}
k\,\tanh(k\,h_{0}) = \frac{\omega^{2}}{g},
\end{equation}
and the vertical mode $w(z)$ is
\begin{equation}
    w(z)=\frac{\cosh\{k\,(z+h_{0})\}}{\cosh(k\,h_{0})}.
\end{equation}
The incident amplitude is $a(\omega)=g\,A\,/\,\omega$, where 
$A=1$\,m is the amplitude of the incident surface elevation,
and the reflected amplitude $b(\omega)$ is found as part of the solution process.

\section{Solution method}\label{sec:method}

Non-dimensional versions of the governing equations given in \S\,\ref{sec:problem} are solved, in which lengths are scaled with respect to the water depth $h_{0}$, and time with respect to $\sqrt{h_0\,/\,g}$. 
The non-dimensional variables are denoted by an overbar, and defined as 
\begin{equation}
    \bar{x}=\frac{x}{h_0},
    \quad
    \bar{z}=\frac{z}{h_0}
    \quad
    \bar{\omega}
    =
    \omega\,\sqrt{h_0\,/\,g}
    \quad
    \bar{\eta}(\bar{x})
    =
    \frac{\eta(x)}{h_0}
    \mathand
    \bar{\phi}(\bar{\mathbf{x}})
    =
    \frac{\phi(\mathbf{x})\,\sqrt{h_{0}\,g}}{h_{0}}.
\end{equation}
The open ocean and cavity domains in non-dimensional variables are, respectively,
\begin{equation}
    \bar{\Omega}_{\textnormal{o}}
    = \{(\bar{x},\bar{z}): \bar{x}<0, -1<\bar{z}<0\}
    \mathand
    \bar{\Omega}_{\textnormal{c}}
    = \{(\bar{x},\bar{z}): 0<\bar{x}<\bar{L}, -\bar{h}(\bar{x})<\bar{z}<-\bar{d}\new{(\bar{x})}\},
\end{equation}
where $\bar{L}=L\,/\,h_{0}$, $\bar{h}(\bar{x})=h(x)\,/\,h_{0}$ and $\bar{d}\new{(\bar{x})}=d\new{(x)}\,/\,h_{0}$.
The governing equations become 
\begin{subequations}
    \label{eq:thinplateok}
    \begin{align}
    \Big(\partial_{\bar{x}}^{2}+\partial_{\bar{z}}^{2}\Big)\,\bar{\phi}=0,
    & \quad\mathbf{\bar{x}}\in\,
    \bar{\Omega}_{\textnormal{o}}
    \cup 
    \bar{\Omega}_{\textnormal{c}},
    \\
    \partial_{\bar{x}}\bar{\phi} =0,
    & \quad \bar{x} = 0, \; -\bar{d}\new{(0)}<\bar{z}<0,
    \\
    \partial_{\bar{x}}\bar{\phi} =0,
    & \quad \bar{x} = \bar{L}, \; -\bar{h}(\bar{L})<\bar{z}<-\bar{d}\new{(\bar{L})},
    \\
    \partial_{\bar{n}}\bar{\phi} = 0,  &\quad 0<\bar{x}<\bar{L}, \; \bar{z} = -\bar{h}(x),\\
    \partial_{\bar{z}}\bar{\phi} =
    0,  &\quad \bar{x}<0, \; \bar{z} = -1,
    \\
    \partial_{\bar{z}}\bar{\phi} = \bar{\omega}^{2}\,\bar{\phi},
    & \quad \bar{x}<0, \; \bar{z} = 0, 
    \\ \label{eq:thinplateok_g}
    \new{\partial_{\bar{n}}\bar{\phi}\approx}
    \partial_{\bar{z}}\bar{\phi} = -\ci\,\bar{\omega}\,\bar{\eta},
    & \quad 0<\bar{x}<\bar{L}, \; \bar{z} = -\bar{d}\new{(\bar{x})},    
    \\ 
    {\partial_{\bar{x}}^{2}\left(\bar{F}(x)\,\partial_{\bar{x}}^{2}\bar{\eta}\right)}
    -\bar{m}(x)\,\bar{\omega}^{2}\,\bar{\eta}
    = 
    -\bar{\eta}
    +
    \ci\,\bar{\omega}\,\bar{\phi},
    & 
    \quad 0<\bar{x}<\bar{L}, \; \bar{z} = -\bar{d}\new{(\bar{x})},
    \end{align}
\end{subequations}
where
$\bar{F}(x)=D(x)\,/\,(g\,\rho_{\textnormal{w}}\,h_0^{4})$
and 
$\bar{m}(x)=\rho_{\textnormal{i}}\,H(x)\,/\,(\rho_{\textnormal{w}}\,h_0)$
are the non-dimensional rigidity and mass of the shelf, respectively, 
plus the non-dimensional version of radiation condition \eqref{eq:radcond}, in which $\bar{w}(\bar{z})=\cosh\{\bar{k}\,(\bar{z}+1)\}\,/\,\cosh(\bar{k})$ and $\bar{k}\,\tanh(\bar{k})=\bar{\omega}^{2}$.
\new{The approximation in \eqref{eq:thinplateok_g} of the normal derivative by the vertical derivative is made for computational convenience.
For the shelf profiles considered in this study, the normal differs from the vertical by $\approx{}1^{\circ}$ at worst 
(for the severely thickening, relatively short shelf considered in 
\S{}\ref{sec:freemodes} and \S{}\ref{sec:thickeningice}), 
but usually far less.}

Overbars are dropped for clarity in the remainder of \S\ref{sec:method}, 
on the understanding that all quantities are non-dimensional unless stated otherwise.


\subsection{Free modes of vibration} 
\label{sec:freemodes}

\begin{figurecross}
\begin{figure}[h!]
\begin{cross}
        \centering
        \setlength{\figurewidth}{0.85\textwidth} 
        \setlength{\figureheight}{0.5\textwidth} 
        \input{Fig02a_R0.tex} 
        \caption{The three different ice shelf thickness profiles considered in this study: (a)~a uniform thickness $H=50$\,m; (b)~mildly thickening shelf, where $H$ increases linearly from $33.3\dot{3}$\,m at $x=0$ to $66.6\dot{7}$\,m at $x=L$; and (c)~a severely thickening shelf, 
         where $H$ increases linearly from $16.6\dot{7}$\,m at $x=0$ to $83.3\dot{3}$\,m at $x=L$. All three shelf profiles share the same mean thickness and length $L=4$\,km.}
        \label{fig:shelf_profiles}
\end{cross}
\end{figure}
\end{figurecross}

\begin{figure}[h!]
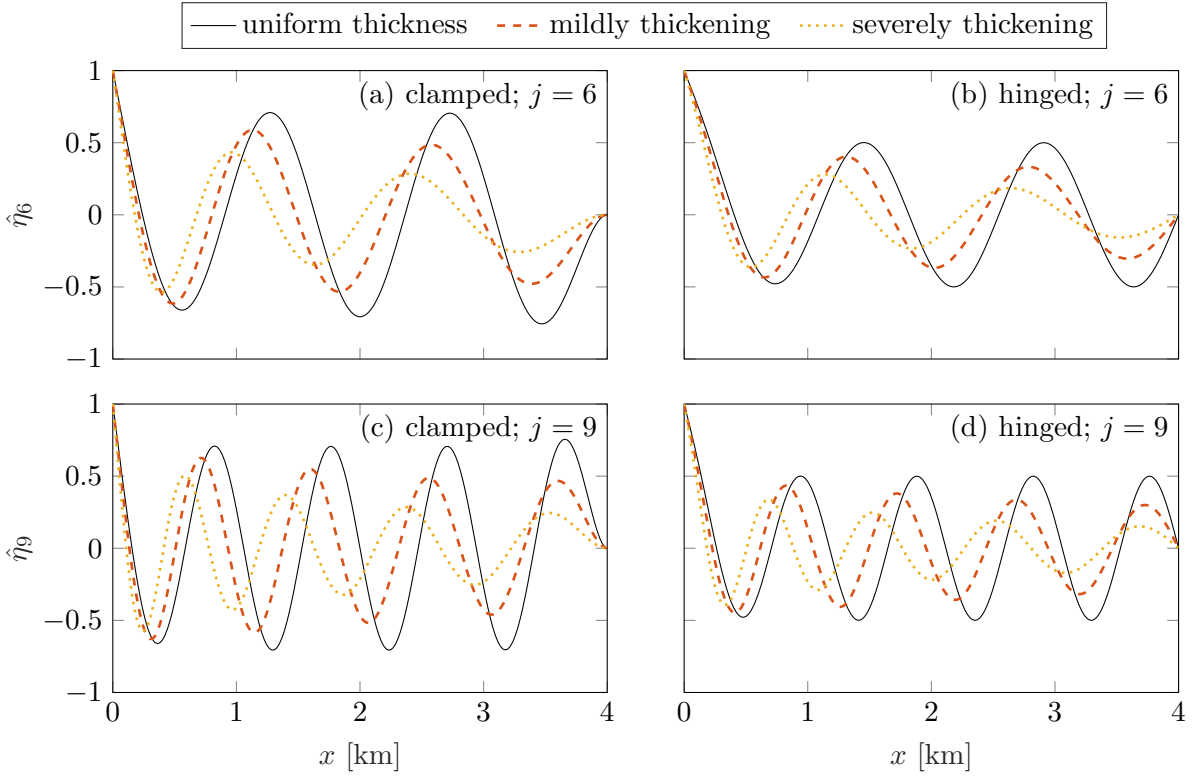

        \centering
        \setlength{\figurewidth}{0.85\textwidth} 
        \setlength{\figureheight}{0.5\textwidth} 
        \begin{cross}
        \input{Fig02_R0.tex} 
        \end{cross}
        \input{Fig02.tex}         
        \caption{Free modes of vibration 
        for mode (a\new{,b})~$j=6$ and (\del{b}\new{c,d})~$j=9$,
        \new{with (a,c)~clamped conditions and (b,d)~hinged conditions at the grounding line,} for \del{the} three shelf profiles\del{ shown in Figure~\ref{fig:shelf_profiles}}, 
        where the thickening-shelf modes are calculated using $N=40$ uniform modes.}
        \label{fig:mode_shapes}
\end{figure}

Free modes of vibration of \del{an} \new{clamped} ice shelf are the eigenfunctions $\eta_{\textnormal{fr}}(x)=\eta_{j}(x)$ ($j=1,2,\ldots$)
of the spectral problem
\begin{subequations}\label{eqs:spectralprb}
\begin{equation}\label{eq:spectral_ODE}
\partial_{x}^{2}
\left(F(x)\,\partial_{x}^{2}\eta_{\textnormal{fr}}\right)
=
\mu\, m(x)\,\eta_{\textnormal{fr}}
\mathfor
0<x<L,
\end{equation}
\begin{equation}\label{eq:free_clamp}
\partial_{x}^{2}\eta_{\textnormal{fr}} =\partial_{x}^{3}\eta_{\textnormal{fr}}=0
\mathfor
x=0
\mathand
\eta_{\textnormal{fr}}  =\partial_{x}\eta_{\textnormal{fr}}
=0
\mathfor 
x=L,
\end{equation}
\end{subequations}
where $\mu$ is the eigenvalue (spectral parameter).
The spectral problem \eqref{eqs:spectralprb} is equivalent to finding the local stationary points (in this case minima) of 
the functional
\begin{equation}\label{eq:functionalJ}
\mathcal{J}\left[ {\eta}_{\textnormal{fr}}\right] =\frac{1}{2}\int_{0}^{L}\left\{F(x) 
\left( \partial_x^2{\eta}_{\textnormal{fr}} \right) ^{2}-\mu\,m(x)\,{\eta}_{\textnormal{fr}}^{2}\right\}\,\d x.
\end{equation}

For an ice shelf of uniform thickness, 
the modes can be found from integration of \eqref{eq:spectral_ODE} to be
    \begin{align}
    \eta_{\textnormal{fr}}(x)=\xi_{j}(x)
    & = C_{j}\Big\{\left(\cos\left(\zeta_{j}\,L\right)+\cosh\left(\zeta_{j}\,L\right)\right)\,\left(\sin\left(\zeta_{j}\,x\right)-\sinh\left(\zeta_{j}\,x\right)\right) \\[8pt]
    &+\left(\sin\left(\zeta_{j}\,L\right)+\sinh\left(\zeta_{j}\,L\right)\right)\,\left(\cosh\left(\zeta_{j}\,x\right)-\cos\left(\zeta_{j}\,x\right)\right)\Big\}
    \end{align} 
    for $j=1,2,\ldots$,
    where the $\zeta_{j}$ ($j=1,2,\dots$) are the (real) roots $\zeta$ of
    \begin{align}
    \cosh\left(\zeta \, L\right)\,\cos\left(\zeta\, L\right)+1& =0,
    \mathX{such that}
    0<\zeta_{1}<\zeta_{2}<\ldots,
    \end{align}
    and the constants $C_{j}$ are chosen such that
    \[
    \int_{0}^{L} \xi_{j}^2\,\d x = 1
    \mathfor
    j=1,2,\ldots.
    \]
    For non-uniform shelves, the modes are found via the Rayleigh--Ritz method, 
    using the expansion 
    \begin{equation}
    \eta_{\textnormal{fr}}(x)\approx\sum_{j=1}^{N} p_{j}\,\xi_{j}\left( x\right),
    \end{equation}
    for a suitably large value of $N$ ($N=20$--40 for the results in this study).
    The functional \eqref{eq:functionalJ} becomes
    \begin{equation}
    \mathcal{J}\left[ \mathbf{p}\right] =\frac{1}{2}\int_{0}^{L}\left\{F(x) \left(\partial_x^2\sum_{j=1}^{N} p_{j}\,\xi_{j}''\left( x\right)
    \right) ^{2}- \mu\,m(x)\left( \sum_{j=1}^{N} p_{j}\,\xi_{j}\left( x\right) \right) ^{2}\right\}\,\d x,
    \end{equation}
    where $\mathbf{p}=[p_{1},p_{2},\ldots,p_{N}]^{T}$.
    Seeking $\partial{}F/\partial{}p_{j}=0$ for $j=1,\ldots,N$ leads to the eigensystem
    \begin{equation}\label{eq:eigen}
    \mathcal{F}\,\mathbf{p}=\mu\,\mathcal{M}\,\mathbf{p},
    \end{equation}
    where the elements of the $N\times{}N$ square matrices $\mathcal{F}$ and $\mathcal{M}$ are
    \begin{equation}\label{eq:K_and_M}
        \{\mathcal{F}\}_{i,j} =
        \int_{0}^{L}F(x)\,
        \xi_{i}''(x) 
        \,
        \xi_{j}''(x) \,\d x
        \mathand
        \{\mathcal{M}\}_{i,j} 
        =
        \int_{0}^{L}m(x)\, \xi_{i}(x)\,\xi_{j}(x)\,\d x,
    \end{equation}
for $i,j=1,2,\ldots,N$.
The eigenvalues of \eqref{eq:eigen} are denoted $\mu=\mu_{j}$ for $j=1,2,\ldots,N$,
where $\mu_{1}<\mu_{2}<\ldots<\mu_{N}$, 
and $\mu_{j}=F\,\zeta_{j}\,/\,m$ for a uniform shelf.
The corresponding eigenvectors are denoted $\mathbf{p}=\mathbf{p}_{j}=[p_{1,j},p_{2,j},\ldots,p_{N,j}]^{T}$ for $j=1,2,\ldots,N$.

Figure~\ref{fig:mode_shapes} shows \del{the} \new{example} free modes of vibration (eigenfunctions \new{$j=6$ and $j=9$})
\del{for modes $j=6$ and $j=9$}
\new{for shelves of length $L=4$\,km, and clamped (left-hand panels) or hinged (right) grounding line conditions, where the modes for hinged conditions are found in a similar fashion to the modes for clamped conditions}.
\new{The shelf geometries are those studied in \S{}\ref{sec:results},
in which the relatively short length is useful to identify resonant behaviours.
Longer shelves have analogous modes to those shown in Figure~\ref{fig:mode_shapes}.}  
The eigenfunctions are normalised to a unit absolute maximum value,
and denoted $\hat{\eta}_{j}(x)$, so that
$\max_{0<x<L}\vert\hat{\eta}_{j}\vert=1$.
The modal index can be deduced from the number of nodes of the eigenfunction (plus the zero at $x=L$).
Corresponding eigenfunctions are
compared for the three ice shelf profiles\del{ shown in 
figure~\ref{fig:shelf_profiles}}: 
a uniform shelf\new{, $H=50$\,m}; a mildly thickening shelf\new{, where $H$ increases linearly from $33.3\dot{3}$\,m at $x=0$ to $66.6\dot{7}$\,m at $x=L$}; 
and a severely thickening shelf\new{, where $H$ increases linearly from $16.6\dot{7}$\,m at $x=0$ to $83.3\dot{3}$\,m at $x=L$}.
The shelves have the same mean thickness, 
so that the thickening shelves are thinner than the uniform shelf for $x<L\,/\,2$, and thicker for $x>L\,/\,2$. 
The eigenfunctions share the same basic shape, with maxima at the shelf front, $x=0$, followed by flexural-wave oscillations that connect smoothly to the zero at the clamped boundary $x=L$.
For the uniform shelf, the amplitude and wavelength of the oscillations are almost constant along the shelf. 
For the thickening shelves, the amplitude reduces, and the wavelength elongates along the shelf, 
with these behaviours most pronounced for the severely thickening shelf.
\new{The corresponding modes for the clamped and hinged conditions are similar, but with the displacement amplitudes along the shelves larger for the clamped conditions than the hinged conditions, relative to the displacement at the shelf front.}


\subsection{Diffraction and radiation potentials}

The ice shelf displacement is decomposed into its free modes of vibration, using the expression
\begin{equation}\label{eq:displacement_expansion}
    \eta(x)\approx
    \sum_{j=1}^{M}{\lambda}_{j}\,{\eta}_{j}(x),
\end{equation}
where $M$ is chosen suitably large and $M\ll{}N$ ($M=10$ in this study).
The complementary decomposition of the velocity potential is
\begin{equation}\label{eq:potential_expansion}
    \phi(x,z)\approx\phi_{0}(x,z)+\sum_{j=1}^{M}{\lambda}_{j}\,{\phi}_{j}(x,z),
\end{equation}
where $\phi_0$ is the diffraction potential, 
and $\phi_{j}$ ($j=1,2,\ldots,M$) are radiation potentials. 
The potentials $\phi_{j}$ ($j=0,1,\ldots,M$) satisfy the equations
\begin{subequations}\label{eqs:poteqns_a}
    \begin{align}
    \Big(\partial_{x}^{2}+\partial_{z}^{2}\Big)\,{\phi_{j}}(\mathbf{x})=0, 
    & 
    \quad
    \mathbf{{x}}\in
    {\Omega}_{\textnormal{o}}\cup 
    {\Omega}_{\textnormal{c}},
    \\
    \partial_{{x}}{\phi_{j}} =0,
    & \quad {x} = 0, \; -d\new{(0)}<{z}<0
    \\
    \partial_{{x}}{\phi_{j}} =0,
    & \quad {x} = L, \; -h(L)<{z}<-d\new{(L)}
    \\
    \partial_{{z}}{\phi_{j}} =
    0,  
    & \quad {x}<0, \; {z} = -1,
    \\
    \partial_{{n}}{\phi_{j}} =
    0,  
    & \quad 0< {x}<L, \; {z} = -{h}(x),
    \\
    \partial_{{z}}{\phi_{j}} 
    = \omega^{2}\,{\phi_{j}},
    & \quad {x}<{0}, \; {z} = 0. 
    \end{align}
\end{subequations}
The diffraction potential, $\phi_{0}$, is the solution of the scattering problem for a stationary shelf,
so that it satisfies the radiation and kinematic conditions
\begin{subequations}
\begin{align}\label{eq:diffraction_radiationcond}
    \phi_{0}(x,z) 
    \sim 
    \{a(\omega)\,\e^{\ci\,k\,x}
    +
    b_{0}(\omega)\,\e^{-\ci\,k\,x}\}\,w(z),
    &
    \quad
    x\to-\infty,
    \\
    \partial_{{z}}{\phi_0} = 0,
    &
    \quad {0}<{x}<L, \; {z} = -{d}\new{(x)}.
\end{align}
\end{subequations}
The diffraction potentials, $\phi_{j}$ ($j=1,2,\ldots,M$), satisfy
\begin{subequations}
\begin{align}\label{eq:radcond_rad}
    \phi_{j}(x,z) 
    \sim 
    b_{j}(\omega)\,\e^{-\ci\,k\,x}\,w(z),
    &
    \quad
    x\to-\infty,
    \\
    \partial_{{z}}{\phi_{j}} = -\ci\,\omega\,\eta_{j},
    &
    \quad {0}<{x}<L, \; {z} = -{d}\new{(x)},
\end{align}
\end{subequations}
where $\eta_{j}(x)$ are the free modes of vibration.
Therefore, let
\begin{equation}\label{eqn:expansion_phi}
\phi_{j}(x,z)
\approx{}
\sum_{i=1}^{N} 
p_{i,j}
\psi_{i}(x,z)
\mathfor
j=1,2\ldots,M,
\end{equation}
where $\psi_{j}$ ($j=1,2,\ldots,N$) are the radiation potentials corresponding to the uniform shelf modes, 
\ie they satisfy (\ref{eqs:poteqns_a}\textit{a}--\textit{d}),
and
\begin{subequations}
\begin{align}\label{eq:radcond_rad}
    \psi_{j}(x,z) 
    \sim 
    \tilde{b}_{j}(\omega)\,\e^{-\ci\,k\,x}\,w(z),
    &
    \quad x\to-\infty,
    \\
    \partial_{{z}}{\psi_{j}} =  -\ci\,\omega\,\xi_{j},
    &
    \quad {0}<{x}<L, \; {z} = -{d}\new{(x)}.
\end{align}
\end{subequations}

Substituting decompositions~(\ref{eq:displacement_expansion}--\ref{eq:potential_expansion}) into the dynamic shelf--cavity boundary condition,
and making use of \eqref{eq:spectral_ODE},
leads to
\begin{equation}\label{eqn:interfacevar}
        \sum_{j=1}^{M}\Big\{
        m(x)\,\mu_{j}
        -
        m(x)\,\omega^{2}
        +
        1    
        \Big\}\,{\lambda}_{j}\,\eta_{j}(x)
        =
        \mathrm{i}\,\omega
        \left(
        \phi_{0}(x,-d\new{(x)})
        +
        \sum_{j=1}^{M} {\lambda}_{j}\,{\phi}_{j}(x,-d\new{(x)})
        \right) ,  
\end{equation}
for $0<x<L$.
Taking inner products with respect to ${\eta}_{j}$ ($j=1,2,\ldots,N$), 
results in the system
\begin{equation}\label{eqn:matrixfinal}
        \left[ 
        \mathcal{W}\,
        \mathcal{P}\,\mathcal{M}\,\mathcal{P}^{T}
        +
        \mathcal{P}\,\mathcal{P}^{T}
        +
        \ci\,\omega\, \mathcal{P}\,\mathcal{A}\,\mathcal{P}^{T}\right]
        \mbox{\boldmath${\lambda}$}
        =
        \ci\,\omega\, \mathcal{P}\,\mathbf{f}, 
\end{equation}
where $\mathcal{W}$ is a $M\times{}M$ diagonal matrix with entries $\{\mathcal{W}\}_{j,j}=\mu_{j}-\omega^{2}$ ($j=1,2,\ldots,M$),
the $M\times{}N$ matrix $\mathcal{P}$ has entries
$\{\mathcal{P}\}_{i,j}=p_{i,j}$ ($i=1,2,\ldots,M$; $j=1,2,\ldots,N$),
the $N\times{}N$ matrix $\mathcal{M}$ is defined in the second component of \eqref{eq:K_and_M},
the $N\times{}N$ matrix $\mathcal{A}$ and
length $N$ column vector $\mathbf{f}$ have entries
\begin{equation}
\{\mathcal{A}\}_{i,j}
=
\int_{0}^{L}\psi_{i}(x,-d\new{(x)})\,\xi_{j}(x)\, \d x
\mathand
\{\mathbf{f}\}_{j}
=
\int_{0}^{L}\phi_{0}(x,-d\new{(x)})\,\xi_{j}(x)\, \d x
,
\end{equation}
for $i,j=1,2,\ldots,N$.
The entries of $\mathcal{A}$ and $\mathbf{f}$ are independent the shelf rigidity, $F(x)$, and mass, $m(x)$, 
and, therefore, can be used for different shelf profiles.
The system is solved for the length $M$ vector 
$\mbox{\boldmath${\lambda}$}=[\lambda_{1},\lambda_{2},\ldots,\lambda_{M}]^{T}$,
once the potentials $\psi_{j}(x,z)$ ($j=0,1,\ldots,N$) have been calculated.

\subsection{Calculation of potentials}

\subsubsection{Non-local operator}

Consider an arbitrary potential $\psi(x,z)=\psi_{j}(x,z)$ for some $j=0,1,\ldots,N$.
In the open ocean region,
the potential can be expressed as
\begin{equation}
\psi(x,z)  \approx
\psi_{\textnormal{inc}}(x,z)
+
\sum_{j=0}^{K}
\beta_{j}\,\e^{\kappa_{j}\,x}\,W_{j}(z)
\mathfor
\bfx\in
\Omega_{\textnormal{o}},
\label{solutionleft}%
\end{equation}
for suitably large $K$ (order 10 used for presented results),
where $\psi_{\textnormal{inc}}=a\,\e^{\i\,k\,x}\,w(z)$ if $j=0$ and $\psi_{\textnormal{inc}}=0$ otherwise.
The vertical functions $W_{j}(z)$ ($j=0,1,\ldots,K$) form an orthonormal set, 
and are defined as
\[
W_{j}(z)=
\sqrt{\frac{4\,\kappa_{j}}{2\,\kappa_{j}+\sin(2\,\kappa_{j})}}\,
\cos\{\kappa_{j}\,(z+1)\}
\mathfor
j=0,1,\ldots,K.
\]
The wavenumbers $\kappa_{j}$ ($j=0,1,\ldots,K$)
satisfy $\kappa_{j}\,\tan(\kappa_{j})=-\omega^{2}$,
with $\kappa_{0}=-\ci\,k$, so that the corresponding term in the series defines the leftward-propagating (reflected) wave, 
and $\kappa_{1}<\kappa_{2}<\ldots<\kappa_{K}\in \mathbb{R}_{+}$, 
so the terms define motions that decay away from the shelf front.

At the interface between the open water and the cavity regions, 
taken from the cavity limit, \ie $x=0_{+}$,
the potential and its normal derivative are expanded as
\begin{equation}
{\phi}(0_{+},z)  
=
\sum_{j=0}^{K}\gamma_{j}\,V_{j}(z)
\mathand
\partial_{x}{\phi}(0_{+},z) 
=
\sum_{j=0}^{K}\delta_{j}\,V_{j}(z)
\mathfor
-1<z<-d\new{(0)},
\end{equation}
where
\begin{equation}
V_{j}(z)=\sqrt{\frac{4\,\chi_{j}}{2\,(1-d\new{(0)})\,\chi_{j} + \sin\{2\,\chi_{j}\,(1-d\new{(0)})\}}}\,
\cos \{\chi_{j}\,(z+1)\},
\end{equation}
in which $\chi_{0}\in\i\,\mathbb{R}_{-}$
and $\chi_{1}<\chi_{2}<\ldots<\chi_{K} \in\mathbb{R}_{+}$
satisfy $\chi_{j}\,\tan\{\chi_{j}\,(1-d\new{(0)})\}=-\omega^{2}$.

Applying continuity of fluid pressure and normal (horizontal) velocity at the interface, gives 
\begin{subequations}
\begin{align}
\label{eq:pot_match}
[\psi_{\textnormal{inc}}]_{x=0}
+
\sum_{j=0}^{K}
\beta_{j}\,W_{j}(z)
& 
=
\sum_{j=0}^{K}\gamma_{j}\,V_{j}(z)
\mathfor
-1<z<-d\new{(0)},
\\
\label{eq:vel_match}
\mathandR
[\partial_{x}\psi_{\textnormal{inc}}]_{x=0}
-
\sum_{j=0}^{K}
\kappa_{j}\,\beta_{j}\,W_{j}(z)
& 
=
\sum_{j=0}^{K}\delta_{j}\,V_{j}(z)
\mathfor
-1<z<-d\new{(0)}.
\end{align}
\end{subequations}
Taking inner-products of \eqref{eq:pot_match} with respect to $V_{j}(z)$ ($j=0,1,\ldots,K$) over $-1<z<-d\new{(0)}$, and \eqref{eq:vel_match} 
with respect to $W_{j}(z)$ ($j=0,1,\ldots,K$)
over $-1<z<0$, making use of the condition $\partial_{x}\psi=0$ for $x=0$ and $-d\new{(0)}<z<0$,
generates the systems
\begin{equation}\label{eq:systems_interface}
    \mathbf{p}_{1}+\mathcal{V}\,\mathbf{b}
    =
    \mathbf{c}
    \mathand
    \mathbf{p}_{2}-
    \mathcal{K}\,\mathbf{b}
    =
    \mathcal{V}^{T}\mathbf{d},
    \mathXL{respectively.}
\end{equation}
In \eqref{eq:systems_interface},
the length $K+1$ column vectors 
are
\begin{equation}
\{\mathbf{b}\}_{j+1}
= 
\beta_{j},
\quad
\{\mathbf{c}\}_{j+1}
= 
\gamma_{j},
\quad
\{\mathbf{d}\}_{j+1}
= 
\delta_{j},
\end{equation}
\begin{equation} 
\{\mathbf{p}_{1}\}_{j+1}
= 
\int_{-1}^{-d}
[\psi_{\textnormal{inc}}]_{x=0}
\,V_{j}\,\d z
\mathand
\{\mathbf{p}_{2}\}_{j+1}
= 
\int_{-1}^{0}
[\partial_{x}\psi_{\textnormal{inc}}]_{x=0}
\,
W_{j}\,\d z
\end{equation}
for $j=0,1,\ldots,K$,
and $(K+1)\times(K+1)$ matrices
are
\begin{equation}
\mathcal{K}=\diag\{\kappa_{0},\kappa_{1},\ldots,\kappa_{K}\}
\mathand
\{\mathcal{V}\}_{j+1,i+1}=\int_{-1}^{-d}W_{i}\,V_{j}\,\d{}z\mathfor i,j=0,\ldots,K.
\end{equation}
The integrals can be calculated explicitly as the functions involved are trigonometric.
Eliminating the vector of unknown amplitudes in the open ocean, $\mathbf{b}$, from \eqref{eq:systems_interface}, 
leads to the Dirichlet-to-Neumann map
\begin{equation}\label{eq:D2Nmap}
\mathbf{d}
=
-\mathcal{V}^{-T}\,\mathcal{K}\,\mathcal{V}^{-1}\,\mathbf{c}
-
\mathcal{V}^{-T}\,\mathcal{K}\,\mathcal{V}^{-1}\,\left(\mathcal{V}\,\mathcal{K} ^{-1}\,\mathbf{p}_{2}-\mathbf{p}_{1}\right),
\end{equation}
from the amplitudes of the potential, $\mathbf{c}$, to the amplitudes of its normal derivative, $\mathbf{d}$.

The Dirichlet-to-Neumann map \eqref{eq:D2Nmap} is used to separate the problem for 
the potential in the cavity, $\psi(x,z)$ for $\bfx\in\Omega_{\textnormal{c}}$,
from the problem for the potential in the open ocean, $\psi(x,z)$ for $\bfx\in\Omega_{\textnormal{o}}$.
A single boundary condition is applied at the interface, which is written
\begin{equation}
\partial_{x}\psi
=
-\mathcal{Q}\{{\psi}\}
- 
q(z)
\mathfor
x=0
\mathand
-1<z<-d\new{(0)},
\end{equation}
where $\mathcal{Q}$ is the non-local operator, such that
\begin{equation}
\mathcal{Q}\{{\bullet}\}(z)  
= 
\sum_{j=0}^{K} 
\alpha_{j}\,V_{p}(z)
\mathwhere
\alpha_{j}
=
\Big\{\mathcal{V}^{-T}\,\mathcal{K}\,\mathcal{V}^{-1}\,
\mathbf{v}
\Big\}_{j+1}
\end{equation}
\begin{equation}
\textnormal{and}\quad
\{
\mathbf{v}
\}_{j+1}=\int_{-1}^{-d}
\bullet(z)\,V_{j}(z)\,\d{}z
\mathfor
j=0,1,\ldots,K,
\end{equation}
and the $q(z)$ is a forcing function defined by
\begin{equation}
q(z) 
= 
\sum_{j=0}^{K} 
{q}_{j} \, V_{j}(z)
\mathwhere
{q}_{j}
= 
\Big\{\mathcal{V}^{-T}\,\mathcal{K}\,\mathcal{V}^{-1}\,\left(\mathcal{V}\,\mathcal{K} ^{-1}\,\mathbf{p}_{2}-\mathbf{p}_{1}\right)\Big\}_{j+1}
\end{equation}
for $j=0,1,\ldots,K$.


\subsubsection{Motion in sub-shelf water cavity}

    
    The finite element method is used to make the remaining calculation of the the potential $\psi(x,z)$ in the 
    cavity region $\Omega_{\textnormal{c}}$.
    Let $\mathcal{T}$ be a quasi-uniform triangulation of the domain $\Omega_{\textnormal{c}}$,
    with vertices $\mathbf{x}_{j}=(x_j,z_j)$ ($j=1,2,\ldots,J$),
    which approximates the varying seabed to an arbitrary degree of accuracy. 
    Let $H^1(\Omega_c)= \{ u \in L^2(\Omega_c), \nabla u \in [L^2(\Omega_c)]^2\}$, and let
    $\mathcal{S}\subset H^{1}(\Omega_{\textnormal{c}})$ be a linear finite element space defined on the $\mathcal{T}$, where
    \begin{equation}
    \mathcal{S}=\left\{\psi\in C^{0}(\Omega_{\textnormal{c}}):\psi|_{T}\in \mathcal{P}\left(T\right),~T\in\mathcal{T}\right\}
    \mathand
    \mathcal{P}(T)
    =
    \{\textnormal{linear polynomials on $T$}\}.
    \end{equation}
    
    The problem is reduced to finding 
    $\psi \in \mathcal{S}$, 
    which satisfies the weak form of
    Laplace's equation in the cavity, \ie
    \begin{equation}\label{eq:FEMweak}
    \int_{\Omega_{\textnormal{c}}}
    \nabla\psi\cdot\nabla\Psi\,\d \mathbf{x}
    =
    \int_{\partial\Omega_{\textnormal{c}}}
    \Psi \,\partial_n\psi \, \d s
    \mathfor
    \Psi \in \mathcal{S},
    \end{equation}
    in which $\partial\Omega_{\textnormal{c}}$ is the cavity boundary and $s$ is the tangential coordinate.
    Using the cavity boundary conditions,
    the right-hand side of \eqref{eq:FEMweak} is
    \begin{equation}
    \label{eq:FEMboundary}
    \int_{\partial\Omega}\Psi\,\partial_n\psi \, \d s
    =
    \int_{-1}^{-d}
    [\Psi\, 
    (\mathcal{Q}\{\psi\}+q)]_{x=0}
    \d x
    +
    \int_{0}^{L}
    [\Psi\,G\{\psi\}]_{z=-d}\,
    \d x,
    \end{equation}
    where $G\{\psi\}=0$ if $\psi=\psi_{0}$ and 
    $G\{\psi\}=-\i\,\omega\,\xi_{j}$ otherwise.
    
    The elements of $\mathcal{S}$ are expressed in terms of the finite element basis functions $\varphi_{j}$ ($j=1,2,\ldots,J$), and, in particular,
    the potential is expressed as
    \begin{equation}
    \psi(x,z)=\sum_{j=1}^{J} c_{j}\,\varphi_{j}(x,z)
    \mathwhere
    c_{j}=\psi(\mathbf{x}_{j})
    \mathfor
    j=1,2,\ldots,J.
    \end{equation} 
    Substituting into the weak form (\ref{eq:FEMweak})
    results in the system
    \begin{subequations}
    \begin{align}
    &\sum_{j=1}^{J}
    \left(\int_{\Omega_{\textnormal{c}}}
    \nabla\varphi_{j}\cdot\nabla\varphi_{n}
    \,\d \mathbf{x}
    -
    \int_{-1}^{-d}
    [\varphi_{n}\,\mathcal{Q}\{\varphi_{j}\}]_{x=0}\,\d z
    \right)\,c_{j}
    \\
    =&
    \int_{-1}^{-d}[\varphi_{n}\,q(z_{n})]_{x=0}\,\d z
    +
    \int_{0}^{L}
    [\varphi_{n}\,G\{x_{n}\}]_{z=-h(x)}\,\d x
    \mathfor
    n=1,2,\ldots,J.
    \end{align}
    \end{subequations}
    The Cholesky factorisation is used to solve the system for $c_{j}$ ($j=1,2,\ldots,J$), 
    noting that the system is sparse, symmetric and positive definite.
    Since we are using the lowest order 
    finite element approach, the discrete solution converges to the exact solution linearly with respect to the mesh-size in the energy norm.
    For the results presented in \S\ref{sec:results}, elements of size $\mathrm{O}(10^{-6})$ are used, 
    giving order $10^{5}$--$10^{6}$ elements in $\Omega_{\textnormal{c}}$.

\begin{cross}
 \section*{4.~Results}
\end{cross}
 
\section{Resonant ice-shelf responses}\label{sec:results}

\begin{cross}
The vibrational responses of the three ice shelves shown in Fig.~\ref{fig:shelf_profiles} 
(uniform, mildly thickening, severely thickening)
to incident swell waves, with periods $T\in(10\,\textnormal{s},50\,\textnormal{s})$,
and over different seabeds,
are investigated.
The shelf length, $L=4$\,km, is relatively short, but is, nevertheless, long with respect to the incident wavelengths, which range from $\approx{}156$\,m at $T=10$\,s 
to $\approx{}2.1$\,km at $T=50$\,s for the $h_{0}=200$\,m open ocean depth used in the results.
Moreover, the short shelf length keeps the resonances well separated in period space, for ease of presentation.
\end{cross}

\new{The effects of seabed shoaling, 
the grounding-line condition (clamped or hinged), and shelf thickening on resonant ice-shelf vibrational responses 
to incident swell waves, with periods $T\in(10\,\textnormal{s},50\,\textnormal{s})$,
are studied.
The shelf length is set as $L=4$\,km, which is short for an ice shelf, but is long with respect to the incident wavelengths that range from $\approx{}156$\,m at $T=10$\,s 
to $\approx{}2.1$\,km at $T=50$\,s for the $h_{0}=200$\,m open ocean depth used in the results.
The short shelf length keeps the resonances well separated in period space, for ease of presentation.
}

\subsection{Shoaling bed}

\begin{figurecross}
\begin{figure}[h!]
\begin{cross}
        \centering
        \setlength{\figurewidth}{0.85\textwidth} 
        \setlength{\figureheight}{0.75\textwidth} 
        \input{Fig03.tex} 
        \caption{Schematics (not to scale)
        of $H=50$\,m thick uniform shelves over (a)~flat bed, (b) mildly shoaling bed, and (c)~steeply shoaling bed.}
        \label{fig:schematic_beds}
\end{cross}
\end{figure}
\end{figurecross}

\begin{figure}[h!]
\centering
\setlength{\figurewidth}{0.9\textwidth} 
\setlength{\figureheight}{0.3\textwidth} 
\input{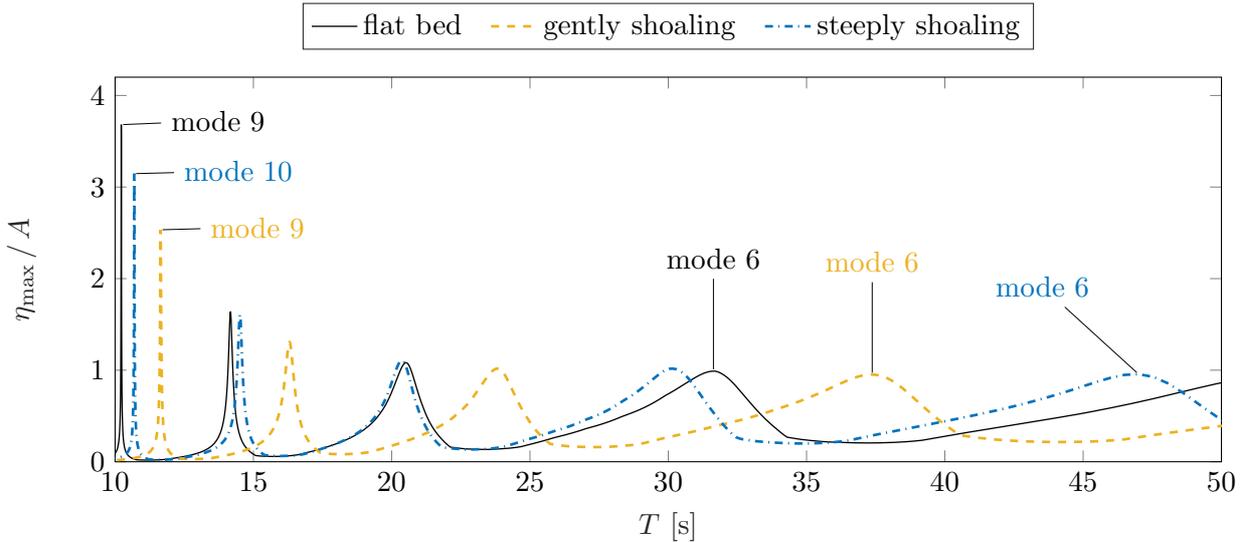} 
\caption{Maximum shelf displacement versus wave period, 
for uniform shelves over \new{shoaling} seabeds\del{ shown in Fig.~\ref{fig:schematic_beds}}.
Corresponding modes of free \del{vibtration} \new{vibration} for shortest- and longest-period resonances are indicated.}
\label{fig:maxdisp_bed}
\end{figure}

\begin{figure}[h!]
\centering
\setlength{\figurewidth}{0.9\textwidth} 
\setlength{\figureheight}{0.5\textwidth} 
\input{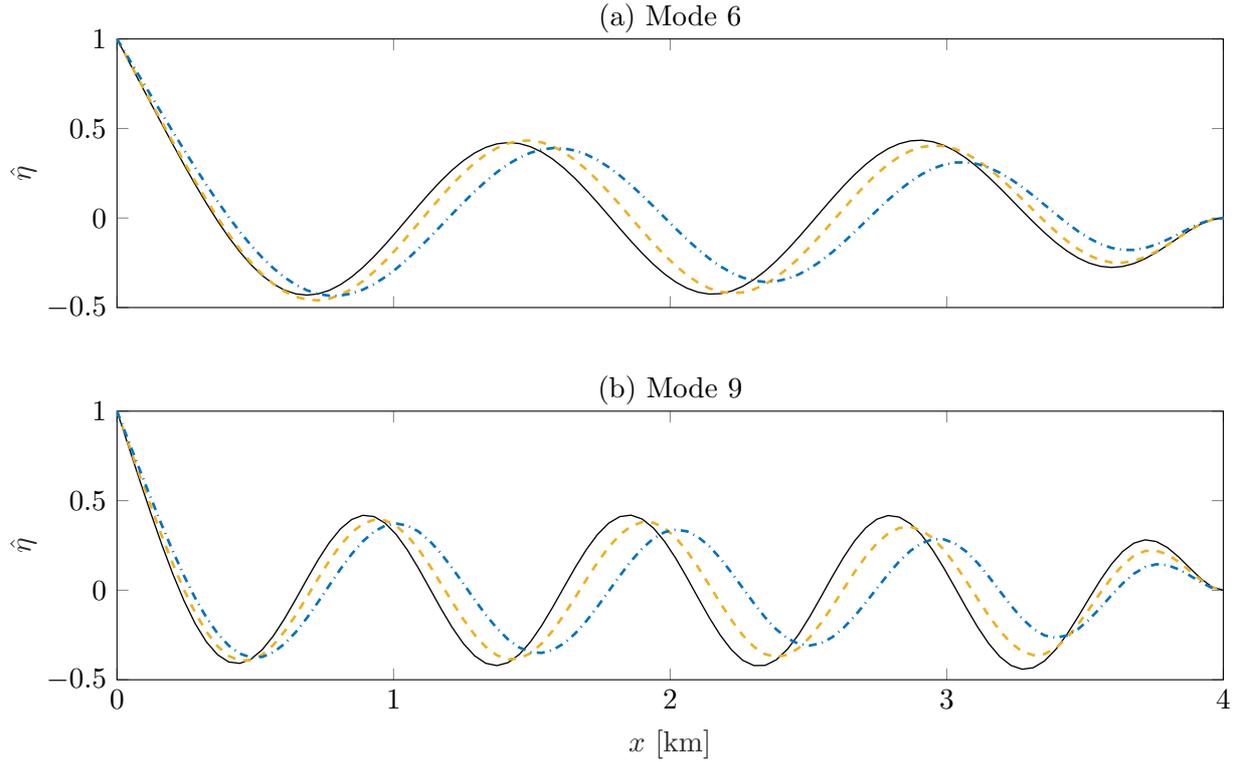} 
\caption{Normalised shelf displacement profiles at resonances corresponding to
(a)~mode~6 and (b)~mode~9,
for the uniform shelves over  \del{seabeds shown in Fig.~\ref{fig:schematic_beds}}  \new{shoaling seabeds}
(legend as in Fig.~\ref{fig:maxdisp_bed}).}
\label{fig:displ_profiles_bed}
\end{figure}



\del{Fig.~\ref{fig:schematic_beds} shows schematics of the three seabed profiles considered, all for the $H=50$\,m thick uniform shelf.}
\new{Three seabeds are considered, all for a $H=50$\,m thick uniform shelf.}
The seabeds differ beneath the shelf ($0<x<L$), in terms of their linear slopes: (a)~the bed is flat beneath the shelf, as in the open ocean (zero slope); 
(b)~the bed gently shoals from $z=-200$\,m at the shelf front, $x=0$, to $z=-120$\,m at the \del{landward end} \new{grounding line}, $x=L$ (2\% slope); and (c)~the bed steeply shoals from $z=-200$\,m at $x=0$ to $z=-40$\,m\del{$(=-d)$} at $x=L$ (4\% slope).
\new{Therefore, the seabeds considered transition between the standard model (zero slope) to a model that eliminates the submarine cliff at the grounding line (steeply sloping), \ie $d(L)=h(L)$.}

The maximum shelf displacement,
\begin{equation}
    \eta_{\textnormal{max}}=\max_{0<x<L}\vert{}\eta\vert
    \left(=\max_{\substack{0<x<L \\ t\in\mathbb{R}}}\vert{}u\vert\right),
\end{equation}
scaled by the incident amplitude, $A$,
is shown versus wave period in Fig.~\ref{fig:maxdisp_bed}, 
for each of the three seabeds\new{ and where the shelf is clamped at the grounding line}.
The responses are dominated by a series of resonances,
which are sharp (narrow-banded) with high peaks for shorter periods,
becoming smoother (broad-banded) with lower peaks as periods increase.
Each resonance is generated by strong excitation of a particular free mode of vibration, and 
the modes associated with the shortest- and longest-period resonances are indicated. 
The modes range from mode~6 (see Fig.~\ref{fig:mode_shapes}a) at the long-period end
to mode~9 (Fig.~\ref{fig:mode_shapes}b; flat bed and gently shoaling) and mode~10 (steeply shoaling) at the short-period end.

The connection between the resonances and the free modes is illustrated by the similarity between the shelf displacement profiles at the resonance peaks (Fig.~\ref{fig:displ_profiles_bed}) and the corresponding free modes (Fig.~\ref{fig:mode_shapes}).
Fluid loading on the shelf from the underlying water cavity causes the resonances to occur at much longer periods than the corresponding free modes. 
For instance, free modes~6 and 9 occur at periods $T=6.43$\,s and $2.69$\,s, respectively, whereas the corresponding peak resonances for the flat seabed occur at $T=32.0$\,s and 10.22\,s, respectively.

Increasing the seabed slope beneath the 
shelf---from flat to mildly shoaling to steeply shoaling---shifts the resonant period 
corresponding to a particular mode to longer periods, 
Hence, it causes the resonance to become smoother and have a lower peak (Fig.~\ref{fig:maxdisp_bed}).
The period shift is greater 
for the lower-order resonances
(\ie at longer periods),
whereas the change in the peak is
more significant for the higher-order resonances 
(shorter periods).
For the flat seabed, 
the amplitudes and lengths of the flexural waves 
are constant along the shelf
(away from the shelf ends; Fig.~\ref{fig:displ_profiles_bed}).
In contrast, 
for the shoaling seabeds,
the amplitudes decrease, and wavelengths increase towards 
the landward end, and
these features are far more pronounced for the steeply shoaling bed
(Fig.~\ref{fig:displ_profiles_bed}).
The behaviour is qualitatively similar to that
of the free modes (Fig.~\ref{fig:mode_shapes})
as the shelf thickening becomes more severe.

\subsection{Grounding-line condition}

\begin{figure}[h!]
\centering
\setlength{\figurewidth}{0.9\textwidth} 
\setlength{\figureheight}{0.6\textwidth} 
\input{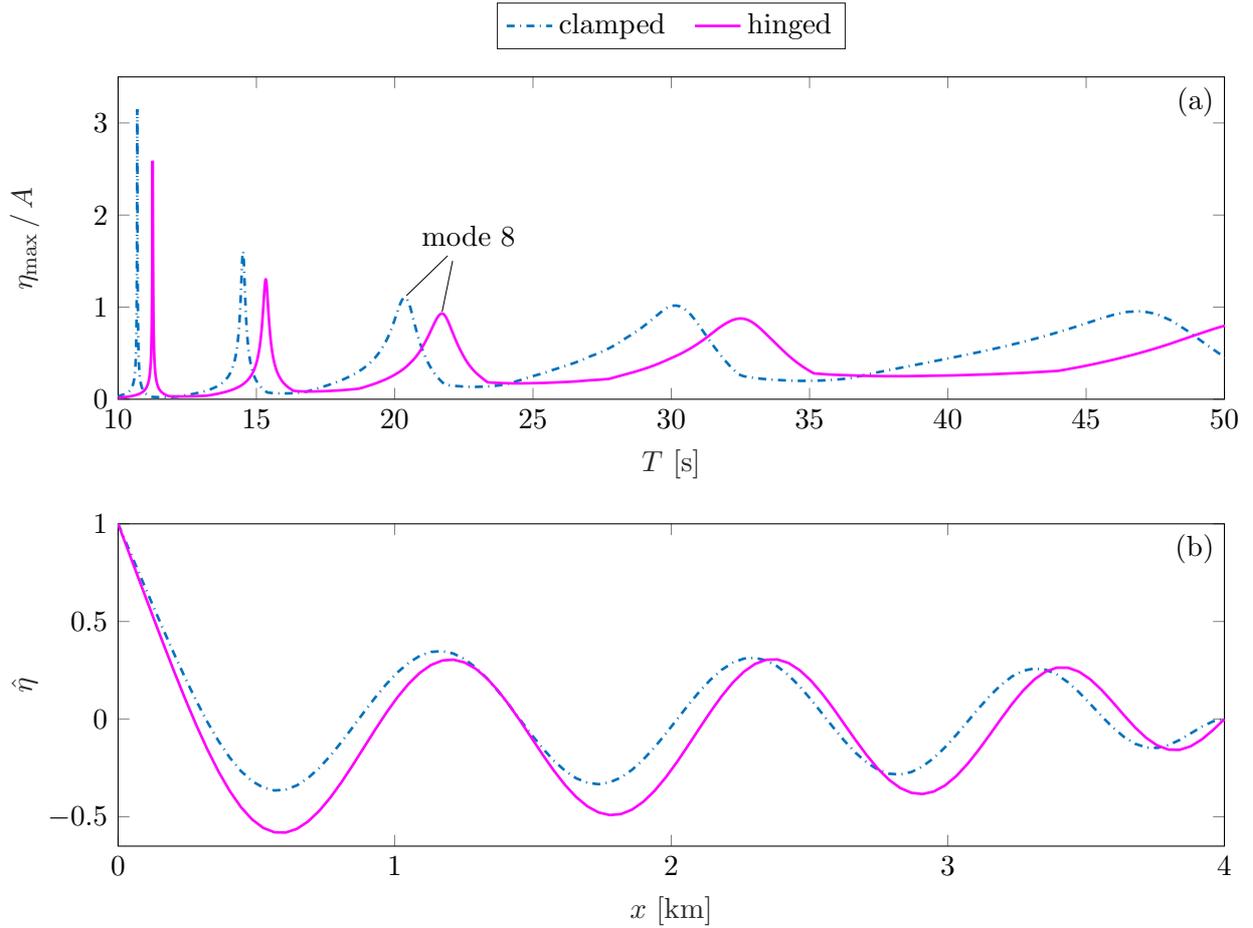} 
\caption{(a)~Maximum shelf displacement versus wave period, 
for uniform shelves over steeply shoal seabeds.
(b)~Normalised shelf displacement profiles at resonances corresponding to mode~8.}
\label{fig:maxdisp_clamphinge}
\end{figure}


\new{Fig.~\ref{fig:maxdisp_clamphinge}a shows the maximum shelf displacement versus wave period for $H=50$\,m thick uniform shelves with clamped or hinged grounding-line conditions, over steeply shoal beds.
The response of the hinged shelf is similar to the clamped shelf, but with the resonances shifted to longer periods, \eg for the mid-range mode~8 resonances indicated in Fig.~\ref{fig:maxdisp_clamphinge}a, 
the peak response occurs at $T=20.4$\,s for the clamped shelf and $T=21.7$\,s for the hinged shelf.
As for the shoaling bed, the period shift is greater at longer periods.
The shift causes the hinged shelf to have slightly smoother resonances with lower peaks than the clamped shelf. 
The corresponding resonant shelf profiles corresponding for the clamped and hinged shelves are also very similar, 
as indicated for mode~8 in Fig.~\ref{fig:maxdisp_clamphinge}b.}




\subsection{Thickening ice}\label{sec:thickeningice}

\begin{figure}[h!]
\centering
\setlength{\figurewidth}{0.9\textwidth} 
\setlength{\figureheight}{0.6\textwidth} 
\input{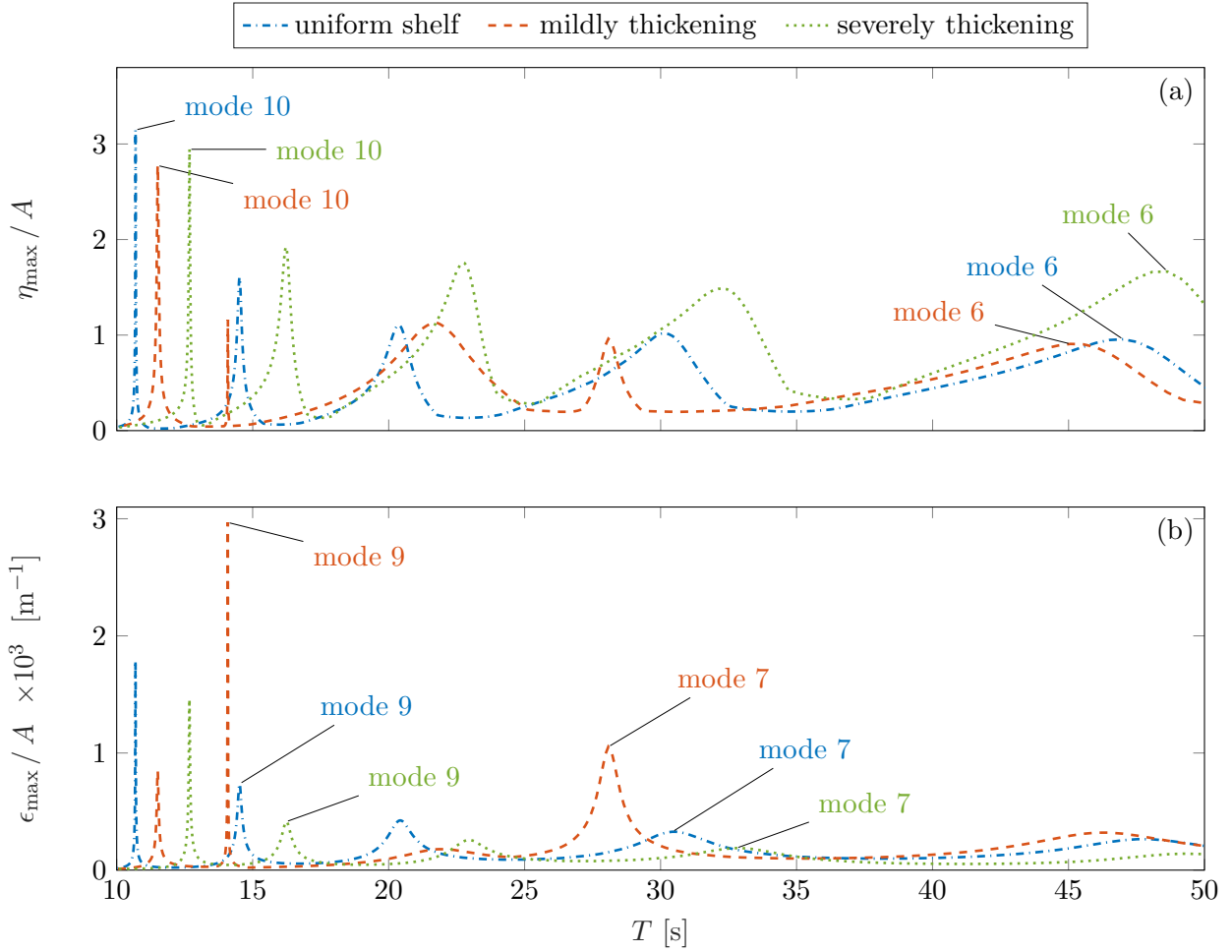} 
\caption{(a)~Maximum displacement and (b)~maximum strain versus wave period for \del{the} three different shelf thickness profiles \del{in Fig.~\ref{fig:shelf_profiles}} over the \del{severely sloping} \new{steeply shoaling} seabed.}
\label{fig:maxdisp_thick}
\end{figure}

\begin{figure}[h!]
\centering
\setlength{\figurewidth}{0.9\textwidth} 
\setlength{\figureheight}{0.5\textwidth} 
\input{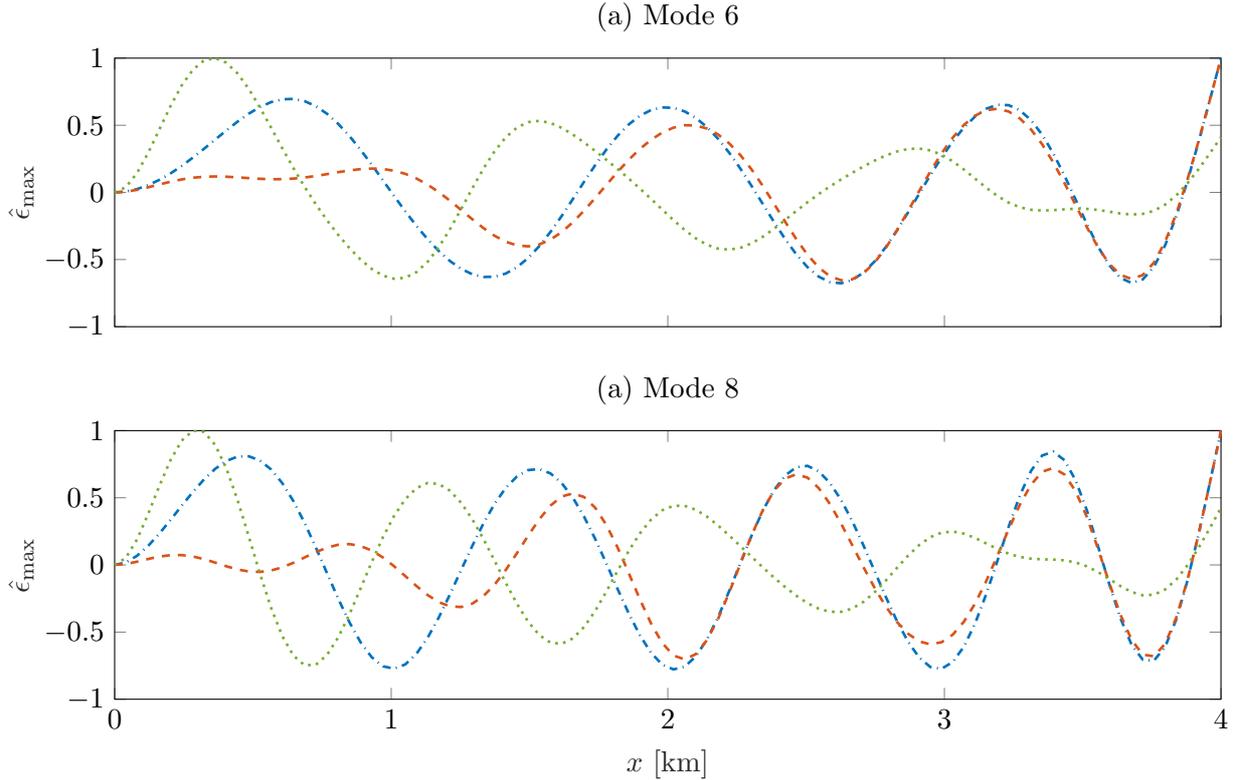} 
\caption{Normalised shelf strain profiles at resonances corresponding to 
(a)~mode~6, and (b)~mode~8,
comparing different shelf thickness profiles (legend as in Fig.~\ref{fig:maxdisp_thick}).}
\label{fig:strain_profiles_thick}
\end{figure}

Fig.~\ref{fig:maxdisp_thick}a shows the maximum shelf displacement versus wave period 
for the \del{three shelves shown in Fig.~\ref{fig:shelf_profiles} (uniform, mildly thickening, severely thickening)}
\new{uniform, mildly thickening, severely thickening shelves (\S{}\ref{sec:freemodes}), which are clamped at the grounding line.}\del{, and
over the steeply shoaling bed (Fig.~\ref{fig:schematic_beds}c).}
\new{\phantom{a} In order to separate the effects of shelf thickening from cavity depth reduction, 
the thickness variations result (artificially) from variations in the upper shelf surface only, with the lower surface constant, $d=40$\,m.
The steeply shoaling seabed  \del{(Fig.~\ref{fig:schematic_beds}c)} is used for each of the shelves, \ie there is no submarine cliff.}
The resonances 
correspond to modes ranging from mode~6 to 10 for all three shelves.
The severely thickening shelf has resonant peaks at the longest periods, 
and the largest peak values (except for the resonance associated to mode~10),
which is presumably due to the maximum displacement occurring at the shelf front (as in Fig.~\ref{fig:displ_profiles_bed}), 
where the severely thickening shelf is thinner than the other two shelves,
which outweighs the tendency of period-lengthening to lower the peaks.
There is no clear relationship between the uniform and mildly thickening shelves' resonances in terms of period shifting or peak values, indicating complexity in the transition of the resonant properties from a uniform shelf to a severely thickening shelf.

The maximum strain,
\begin{equation}
    \epsilon_{\textnormal{max}}
    =
    \max_{0<x<L}
    \vert \epsilon(x)\vert
    \mathwhere
    \epsilon(x)
    =
    \frac{1}{2}\,H(x)\,\eta''(x) 
\end{equation}
is shown versus wave period
in Fig.~\ref{fig:maxdisp_thick}b
for the three shelves.
Resonant peaks are evident at the same periods as the maximum displacements (Fig.~\ref{fig:maxdisp_thick}a). 
However, the relative sizes of the resonant peaks between the shelves are markedly different for the strains compared to the displacements.
The mode~6 and mode~8 peaks are prominent,
and the peak values for the mildly thickening shelf far outweigh those of the uniform and severely thickening shelves, despite the mildly thickening shelf having the smallest peak values for the corresponding displacement peaks.

The maximum strains at the resonant peaks occur at the \del{landward shelf end} \new{grounding line}
for the uniform and mildly sloping shelves (Fig.~\ref{fig:strain_profiles_thick}). 
For the uniform shelf, 
the amplitude of the strain oscillations is nearly constant along the shelf, 
and the maximum value at the 
\del{landward end} \new{grounding line} only slightly exceeds the peaks along the shelf.
For the mildly thickening shelf, the amplitude of the oscillations grows along the shelf, so that the maximum at the \del{landward end}
\new{grounding line} far exceeds the strains towards shelf front. 
Therefore, the increase in strain $\epsilon(x)$ along the shelf due to $H(x)$ increasing 
dominates the strain reduction due to $\eta''(x)$ decreasing along the shelf 
(which can be inferred from the decreasing amplitudes and increasing wavelengths in the displacement profiles; Fig.\ \ref{fig:mode_shapes}).
For the severely thickening shelf, 
$\eta''$ decreasing outweighs $H$ increasing, 
and the maximum strain is attained in the shelf interior close to the shelf front.


\begin{cross}
\subsection{Viscoelastic damping}

The persistence of the resonances to viscoelastic damping of the shelf vibrations is tested by mapping the real-valued Young's modulus
to a complex-valued, frequency-dependent Young's modulus 
\begin{equation}
    E\mapsto{}
    E\,(1-\ci\,\omega\,\sigma)
    \quad\textnormal{where $\sigma$ is a damping parameter,}
\end{equation}
using the Kelvin--Voigt model.
The value $\sigma=0.1$ effectively removes the highest-order resonances (mode~10),
but $\sigma=1$ is required to remove lower-order resonances (Fig.\ \ref{fig:displ_vs_T_damp}), 
and is not large enough to completely remove the lowest-order resonances (modes~6--7).
The damping has a greater impact on the maximum displacements of the uniform shelf 
(Fig.\ \ref{fig:displ_vs_T_damp}a) than the 
severely thickening shelf (Fig.\ \ref{fig:displ_vs_T_damp}b), 
because the resonant peaks for the severely thickening shelf are
shifted to longer periods.
\end{cross}

\begin{figurecross}
\begin{figure}[h!]
	\centering
	\setlength{\figurewidth}{0.9\textwidth} 
	\setlength{\figureheight}{0.6\textwidth} 
	\input{Fig09.tex} 
	\caption{Normalised maximum displacement versus wave period, comparing different damping parameter values, for (a)~uniform shelf and (b)~severely thickening shelf, over the steeply shoaling seabed.}
	\label{fig:displ_vs_T_damp}
\end{figure}
\end{figurecross}

\begin{cross}
\subsection{Irregular wave fields and transfer function}

Inspired by recently reported measurements of ice shelf vibrations and concomitant open ocean wave activity at a small number of points  in the vicinity of the ice shelf front \cite{chen2019ross},
the model is used to simulate time series at a nominal 
sample point in the ocean ($x=-10$\,km)  and on the ice shelf ($x=10$\,km), similar to \cite{chen2019ross}.
Irregular incident wave fields are prescribed, 
based on the Pierson--Moskowitz (fully developed) spectrum \cite{Kahma2005}
\begin{equation}
     S_{\textnormal{in}}(\omega: T_{\textnormal{p}}) 
     = 
     \frac{\alpha\,g^2}{\omega^5}\exp\left(-\beta\left(\frac{\omega_0}{\omega}\right)^4\right),
\end{equation}
where $ \alpha = 8.1\times 10^{-3}$, $ \beta = 0.74 $, $\omega_0 = g\, / \,U_{19.5}$ and $U_{19.5}$ is the wind speed at 19.5\,m above the sea surface,
which is chosen to give a prescribed peak period
\begin{equation}
    T_{\textnormal{p,in}} 
    = \frac{2\,\pi}{\omega_{\textnormal{p,in}}}
    \quad\textnormal{where}\quad
    \omega_{\textnormal{p,in}} = \frac{0.877 \, g}{U_{19.5}}.
\end{equation}
Figure~\ref{fig:timeseries} shows time series of 
randomly generated instances (random phases) of the incident wave field at peak period $T_{\textnormal{p,in}}=20$\,s, along with corresponding time series 
of the ice shelf vibrations. 
The displacements are scaled by the significant wave height of the incident field, $H_{\textnormal{s,in}}$,
where
\begin{equation}
    H_{\textnormal{s,in}} = 4 \sqrt{
    \,\int_{0}^{\infty}S_{\textnormal{in}}(\omega)\wrt\omega.}
\end{equation}
Results are given for the uniform shelf and the severely thickening shelf (both over a steeply shoaling seabed),
and for zero damping ($\sigma=0$) and the largest damping parameter considered ($\sigma=1$).
For zero damping, the ice shelves experience large displacements, 
of up to 8--9\% of the significant wave height, $H_{\textnormal{s,in}}$,
and the displacement magnitudes for the uniform and severely sloping shelves are comparable at the measurement location.
The large damping value, $\sigma$, reduces the displacements, but 
the maximum values are still 3--4\% of $H_{\textnormal{s,in}}$.
\end{cross}

\begin{cross}
An ocean-to-shelf transfer function for displacement is defined as
\begin{equation}
    \mathcal{T}_{\textnormal{d}}(x)=\frac{H_{\textnormal{s,sh}}(x)}{H_{\textnormal{s,in}}}
    \mathwhere
    H_{\textnormal{s,sh}}(x) =
    4\,\sqrt{\,\int_{0}^{\infty}
    S_{\textnormal{in}}(\omega)\,
    \left|\eta(x)\right|^{2}
    \wrt\omega},
\end{equation}
is the significant displacement height in the shelf at location $x$.
The transfer function increases monotonically with increasing peak period (Fig.~\ref{fig:transfer}a). 
At $x=10$\,km, the displacement is marginally larger for the uniform shelf than the severely sloping shelf (by mean factor 1.15) for incident peak periods up to $T_{\textnormal{p,in}}\approx{}22$\,s, as 
the amplitude decrease along the shelf is greater for severely thickening shelf, overcoming the larger maximum displacements at the shelf front for the severely thickening shelf (Fig.~\ref{fig:maxdisp_thick}a). 
In contrast, for peak periods $T_{\textnormal{p,in}}>{}22$\,s,
the amplitude decrease is not strong enough to 
overcome larger amplitudes of the severely thickening shelf at $x=10$\,km,
so that
the transfer function is greater for the severely thickening shelf than the uniform shelf,
and the the difference between the transfer functions becomes greater as incident peak period increases, up to a factor 1.52 at $T_{\textnormal{p,in}}=30$\,s.
At $x=10$\,km, the transfer function is in the range 0.02 to 0.14.
Results are also shown for the transfer function at the shelf front ($x=0$), 
where the values are much greater, ranging from 
0.08 at $T_{\textnormal{p,in}}=10$\,s for the uniform shelf,
to 0.74 for $T_{\textnormal{p,in}}=30$\,s for the severely thickening shelf.
At the shelf front, the displacement of the severely thickening shelf is greater than the uniform shelf for all incident peak periods in the chosen range,
with the difference increasing as peak period increases, up to factor 1.78 at  $T_{\textnormal{p,in}}=30$\,s.
\end{cross}

\begin{cross}
In analogy with the displacement transfer function, a transfer function for strain is defined as
\begin{equation}
    \mathcal{T}_{\textnormal{s}}(x)=
    \frac{H_{\textnormal{s},\textnormal{sh}}(x)}{H_{\textnormal{s,in}}}
    \mathwhere
    H_{\textnormal{s},\textnormal{sh}}(x) =
    4\,\sqrt{\,\int_{0}^{\infty}
    S_{\textnormal{in}}(\omega)\,
    \left|\epsilon(x)\right|^{2}
    \wrt\omega}
\end{equation}
is a significant strain height \cite{williams2013wavea,williams2013waveb}.
Figure~\ref{fig:transfer}b shows the strain transfer function for the uniform and severely thickening shelves at $x=10$\,km 
(the strain is zero at the shelf front).
Similarly to the displacement transfer function at $x=10$\,km,
the strain transfer function is greater for the uniform shelf for shorter incident peak periods (up to $T_{\textnormal{p,in}}\approx{}24$\,s),
and for the severely thickening shelf for longer peak periods ($T_{\textnormal{p,in}}>24$\,s),
with the difference increasing as incident peak period
increases, up to a factor 1.33 at $T_{\textnormal{p,in}}=30$\,s.
\end{cross}

\begin{figurecross}
\begin{figure}[h!]
\centering
\setlength{\figurewidth}{0.9\textwidth} 
\setlength{\figureheight}{0.6\textwidth} 
\input{Fig11.tex} 
\caption{Normalised surface displacement time series 
at $x=-10$\,km for irregular incident wave field with peak period $T_{\textnormal{p,in}}=20$\,s, and corresponding shelf displacement at $x=10$\,km.}
\label{fig:timeseries}
\end{figure}
\end{figurecross}




\begin{figurecross}
\begin{figure}[h!]
\centering
\setlength{\figurewidth}{0.9\textwidth} 
\setlength{\figureheight}{0.6\textwidth} 
%
\definecolor{mycolor1}{rgb}{0.00000,0.44700,0.74100}%
\definecolor{mycolor2}{rgb}{0.92900,0.69400,0.12500}%
\begin{tikzpicture}

\begin{axis}[%
name=myaxis_top,
width=0.951\figurewidth,
height=0.48\figureheight,
at={(0\figurewidth,0.581\figureheight)},
scale only axis,
xmin=10,
xmax=30,
xtick={10,15,20,25,30},
xticklabels={\empty},
ymin=0,
ymax=1.1,
ytick={  0, 0.2, 0.4, 0.6, 0.8,   1},
ylabel style={font=\color{white!15!black}},
ylabel={$\mathcal{T}_{\textnormal{d}}$},
axis background/.style={fill=white},
legend style={at={(axis cs: 10.25,1.075)}, anchor=north west, legend cell align=left, align=left, draw=white!15!black}
]
\addplot [color=mycolor1, line width=1.0pt]
  table[row sep=crcr]{%
10	0.0244812362072828\\
11	0.0348290459263154\\
12	0.0424999333980622\\
13	0.0472424635297877\\
15	0.0539571592309578\\
16	0.058511781310596\\
18	0.0691543519606697\\
19	0.0738691498586377\\
20	0.0776534332379875\\
21	0.080428811110778\\
22	0.0822770792505203\\
23	0.0833779887538668\\
25	0.0842845162674095\\
27	0.0850759977664914\\
28	0.0859527639837658\\
29	0.0873616531479264\\
30	0.0894047399055928\\
};
\addlegendentry{$x=10\,\textnormal{km}$ uniform shelf}

\addplot [color=mycolor2, line width=1.0pt]
  table[row sep=crcr]{%
10	0.026412232998041\\
12	0.0311998650423817\\
13	0.0343145524293114\\
14	0.0385650838466844\\
15	0.044011121979846\\
16	0.0504428297410087\\
17	0.0572108411091783\\
18	0.063523610825321\\
19	0.0688798755234963\\
20	0.0732791026855395\\
22	0.0813679924799864\\
23	0.0864607408234583\\
24	0.0927652907388925\\
25	0.100115688859209\\
27	0.115949427065392\\
28	0.12336569723627\\
29	0.129930352074808\\
30	0.135446721506348\\
};
\addlegendentry{$x=10\,\textnormal{km}$ severely thickening}

\addplot [color=mycolor1, dashdotted, line width=1.0pt]
  table[row sep=crcr]{%
10	0.0781872576635294\\
11	0.103804331376075\\
12	0.122732188796789\\
13	0.138722741654018\\
14	0.158321736485824\\
15	0.183138838820994\\
16	0.209303579437179\\
17	0.232623584037512\\
18	0.251556844549778\\
19	0.267124835445376\\
20	0.28160487168952\\
21	0.297020388115719\\
22	0.314125106795121\\
24	0.350283705131933\\
25	0.366695190543421\\
26	0.380679858777011\\
27	0.391929702570049\\
28	0.400635653293129\\
29	0.407340865701521\\
30	0.412775393630831\\
};
\addlegendentry{$x=0$ uniform shelf}

\addplot [color=mycolor2, dashdotted, line width=1.0pt]
  table[row sep=crcr]{%
10	0.15285107721656\\
11	0.186301539734345\\
12	0.228277464569835\\
13	0.267616861476274\\
14	0.302187352961539\\
15	0.337635595357355\\
16	0.378645168790509\\
17	0.423752961762961\\
18	0.467879063486937\\
19	0.506965475199603\\
20	0.539905838688131\\
21	0.567968367666328\\
22	0.593278570663951\\
23	0.617464878846402\\
24	0.641031397255645\\
25	0.663495305364716\\
26	0.683932111846794\\
27	0.701519194314766\\
28	0.715863216520304\\
29	0.727094399937432\\
30	0.735796399688301\\
};
\addlegendentry{$x=0$ severely thickening}

\node[below left]
at (axis cs: 30,1.1) {(a)~displacement};

\end{axis}

\begin{axis}[%
name=myaxis_bot,
width=0.951\figurewidth,
height=0.48\figureheight,
at={(0\figurewidth,0\figureheight)},
scale only axis,
xmin=10,
xmax=30,
xtick={10, 15, 20, 25, 30},
xlabel style={font=\color{white!15!black}},
xlabel={$T_{\textnormal{p,in}}$ [s]},
ymin=0,
ymax=0.8,
ytick={  0, 0.2, 0.4, 0.6, 0.8},
ylabel style={font=\color{white!15!black}},
ylabel={$\mathcal{T}_{\textnormal{s}}$~$\times{}10^{5}$~~[m$^{-1}$]},
axis background/.style={fill=white},
legend style={at={(axis cs: 29.75,0.025)}, anchor=south east, legend cell align=left, align=left, draw=white!15!black}
]
\node[below right]
at (axis cs: 10,0.8) {(b)~strain};

\addplot [color=mycolor1, line width=1.0pt]
  table[row sep=crcr]{%
10	0.222363541948567\\
11	0.297805952604961\\
12	0.365855073473043\\
13	0.411750498526843\\
14	0.439995663825389\\
15	0.461633766591365\\
16	0.48435413318786\\
17	0.509287556092453\\
18	0.533655216364586\\
19	0.554292721190343\\
20	0.569287053493778\\
21	0.578111411778188\\
22	0.581242525847149\\
23	0.57974227651755\\
24	0.574925187870825\\
25	0.568123085262521\\
27	0.553178411754015\\
28	0.546828172356818\\
29	0.542068638531262\\
30	0.539284020730484\\
};
\addlegendentry{$x=10\,\textnormal{km}$ uniform shelf}

\addplot [color=mycolor2, line width=1.0pt]
  table[row sep=crcr]{%
10	0.22815906147763\\
11	0.269384385967765\\
12	0.302750097368516\\
13	0.336644937760166\\
14	0.371400595367877\\
15	0.404707900816994\\
16	0.434688991011349\\
17	0.459860482647994\\
18	0.47923975516969\\
19	0.492978335891845\\
20	0.502752660829259\\
21	0.511561543249709\\
22	0.522874821012124\\
23	0.539364933009768\\
24	0.561879413365777\\
25	0.589292990319059\\
27	0.64894807426095\\
28	0.676215116830875\\
29	0.699460398048036\\
30	0.71788849866379\\
};
\addlegendentry{$x=10\,\textnormal{km}$ severely thickening}

\end{axis}

\draw[red, line width=2pt](myaxis_top.outer north west)--(myaxis_bot.outer south east); 
\draw[red, line width=2pt](myaxis_top.outer north east)--(myaxis_bot.outer south west); 

\end{tikzpicture}%
\caption{Transfer functions versus peak period,
for (a)~displacement and (b)~strain,
with damping parameter $\sigma=0.1$.}
\label{fig:transfer}
\end{figure}
\end{figurecross}


\section{\new{Realistic shelf--cavity geometries and irregular waves fields}}

\begin{figure}[h!]
\centering
\setlength{\figurewidth}{0.9\textwidth} 
\setlength{\figureheight}{0.6\textwidth} 
\input{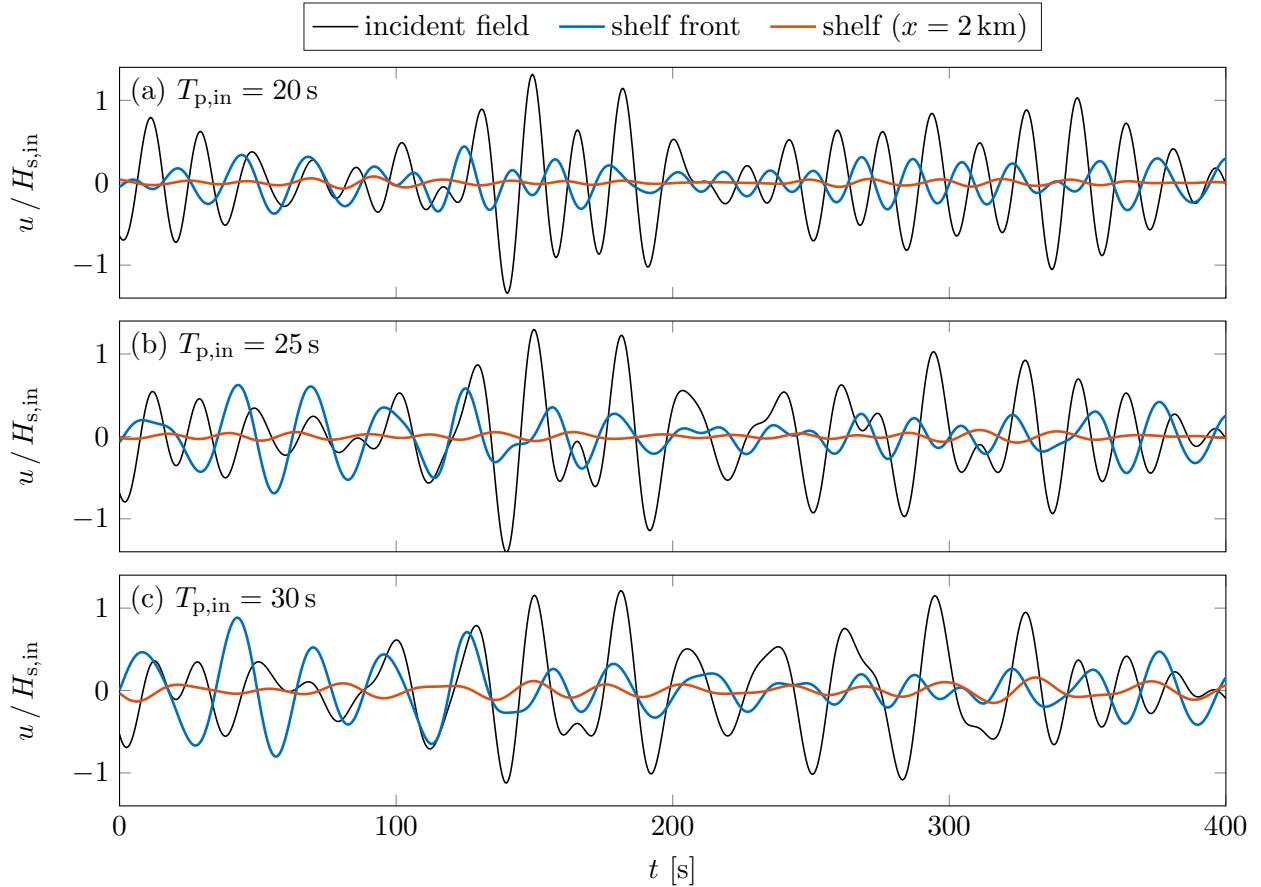} 
\caption{Normalised surface displacement time series for 50\,km-long severely-thickening shelf over steeply shoaling bed
forced by irregular incident wave field with peak period $T_{\textnormal{p,in}}$, showing
incident displacements in the ocean ($x=-2$\,km), and shelf displacements at the shelf front and $x=2$\,km onto shelf.}
\label{fig:timeseries_new}
\end{figure}

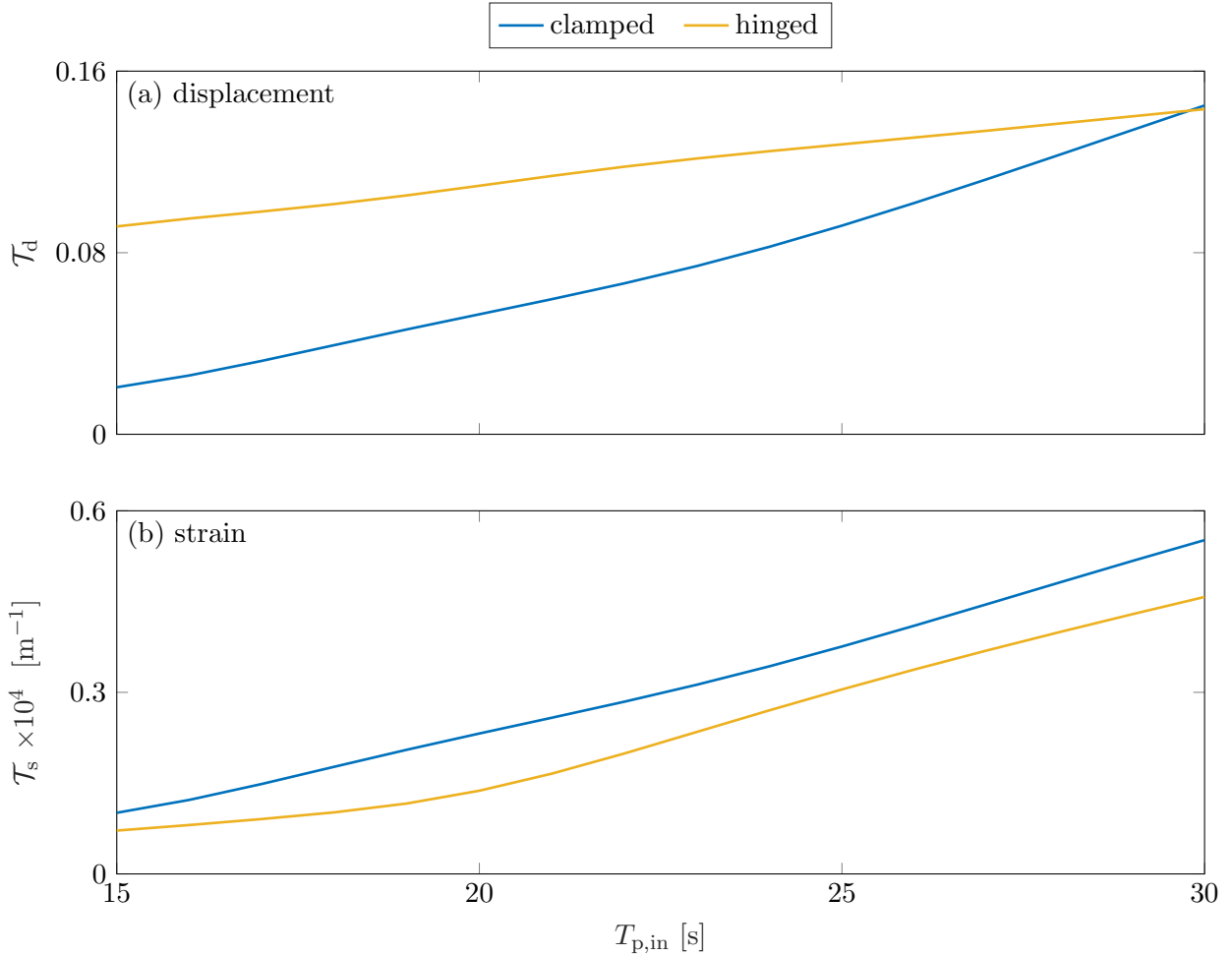
\begin{figure}[h!]
\centering
\setlength{\figurewidth}{0.9\textwidth} 
\setlength{\figureheight}{0.6\textwidth} 
%
\definecolor{mycolor1}{rgb}{0.00000,0.44700,0.74100}%
\definecolor{mycolor2}{rgb}{0.92900,0.69400,0.12500}%
\begin{tikzpicture}

\begin{axis}[%
name=myaxis_top,
width=0.951\figurewidth,
height=0.48\figureheight,
at={(0\figurewidth,0.581\figureheight)},
scale only axis,
xmin=15,
xmax=30,
xtick={15,20,25,30},
xticklabels={\empty},
ymin=0,
ymax=0.16,
ytick={0, 0.08, 0.16},
ylabel style={font=\color{white!15!black}},
yticklabel style={
        /pgf/number format/fixed,
        /pgf/number format/precision=5
},
ylabel={$\mathcal{T}_{\textnormal{d}}$},
axis background/.style={fill=white},
legend style={
at={(axis cs:22.5,0.17)},
legend columns=-1,
/tikz/every even column/.append style={column sep=3mm},
anchor=south,
draw=white!15!black}
]
\node[below right]
at (axis cs: 15,0.16) {(a)~displacement};
\addplot [color=mycolor1, line width=1.0pt]
  table[row sep=crcr]{%
15	0.0207092408258873\\
16	0.0259014928817223\\
17	0.0323316759245849\\
19	0.0461906906508638\\
21	0.0595249557401054\\
22	0.0664957366376946\\
23	0.0741247941863215\\
24	0.0825995656099892\\
25	0.0919132665580413\\
26	0.101911724049753\\
27	0.112385836594598\\
28	0.123139919040742\\
30	0.144852500543774\\
};
\addlegendentry{clamped}

\addplot [color=mycolor2, line width=1.0pt]
  table[row sep=crcr]{%
15	0.0915705706143228\\
16	0.0950609021133708\\
17	0.0981444174898094\\
18	0.101433006601454\\
19	0.105247853730564\\
20	0.109497133043856\\
21	0.113829188598391\\
22	0.11789259254741\\
23	0.121513902590927\\
24	0.124731830059041\\
27	0.133749941536237\\
30	0.14324119538783\\
};
\addlegendentry{hinged}

\end{axis}

\begin{axis}[%
name=myaxis_bot,
width=0.951\figurewidth,
height=0.48\figureheight,
at={(0\figurewidth,0\figureheight)},
scale only axis,
xmin=15,
xmax=30,
xtick={15, 20, 25, 30},
xlabel style={font=\color{white!15!black}},
xlabel={$T_{\textnormal{p,in}}$ [s]},
ymin=0,
ymax=0.6,
ytick={0, 0.3, 0.6},
ylabel style={font=\color{white!15!black}},
ylabel={$\mathcal{T}_{\textnormal{s}}$~$\times{}10^{4}$~~[m$^{-1}$]},
axis background/.style={fill=white},
legend style={at={(axis cs: 29.75,0.025)}, anchor=south east, legend cell align=left, align=left, draw=white!15!black}
]
\node[below right]
at (axis cs: 15,0.6) {(b)~strain};
\addplot [color=mycolor1, line width=1.0pt]
  table[row sep=crcr]{%
15	0.100999004948964\\
16	0.122294607832405\\
17	0.148616244062911\\
18	0.177052090397741\\
19	0.205160116580714\\
20	0.232002366315378\\
21	0.258068562260842\\
22	0.28455793184926\\
23	0.312650807584884\\
24	0.343033549468981\\
25	0.375724637887391\\
26	0.410204847559349\\
27	0.445715709944828\\
28	0.481511932537156\\
29	0.516965028013711\\
30	0.551560362004057\\
};

\addplot [color=mycolor2, line width=1.0pt]
  table[row sep=crcr]{%
15	0.0715862057922081\\
16	0.0807498403052982\\
17	0.0906763753924551\\
18	0.101698650011734\\
19	0.116326479734745\\
20	0.137469625448311\\
21	0.165678655252993\\
22	0.198971902485443\\
24	0.270298343879553\\
25	0.304826308503266\\
26	0.337755627092488\\
27	0.369218815911797\\
28	0.39954305379505\\
29	0.428987211884177\\
30	0.457621610110021\\
};

\end{axis}
\end{tikzpicture}%
\caption{Transfer functions versus incident peak period,
for (a)~displacement and (b)~strain,
taken at $x=2$\,km.}
\label{fig:transfer_new}
\end{figure}

\new{
Inspired by recently reported measurements of ice shelf vibrations and concomitant open ocean wave activity at a small number of points  in the vicinity of the ice shelf front \cite{chen2019ross},
the model is used to simulate time series at a nominal 
sample point in the ocean ($x=-2$\,km) and on the ice shelf ($x=2$\,km), similar to \cite{chen2019ross}.
Irregular incident wave fields are prescribed, 
based on the Pierson--Moskowitz (fully developed) spectrum \cite{Kahma2005}
\begin{equation}
     S_{\textnormal{in}}(\omega: T_{\textnormal{p}}) 
     = 
     \frac{\alpha\,g^2}{\omega^5}\exp\left(-\beta\left(\frac{\omega_0}{\omega}\right)^4\right),
\end{equation}
where $ \alpha = 8.1\times 10^{-3}$, $ \beta = 0.74 $, $\omega_0 = g\, / \,U_{19.5}$ and $U_{19.5}$ is the wind speed at 19.5\,m above the sea surface,
which is chosen to give a prescribed peak period
\begin{equation}
    T_{\textnormal{p,in}} 
    = \frac{2\,\pi}{\omega_{\textnormal{p,in}}}
    \quad\textnormal{where}\quad
    \omega_{\textnormal{p,in}} = \frac{0.877 \, g}{U_{19.5}}.
\end{equation}
A $L=50$\,km long shelf is considered, which linearly thickens from $H(0)=16.6\dot{7}$\,m at the shelf front to $H(L)=83.3\dot{3}$\,m at the grounding line (same as the severely thickening shelves considered in \S\ref{sec:freemodes} and \S\ref{sec:results}),
with $d(x)=(\rho_{\textnormal{i}}\,/\,\rho_{\textnormal{w}})\,H(x)\approx0.9\,H(x)$. 
The seabed beneath the shelf linearly shoals from $h(0)=h_{0}=200$\,m at the shelf front to $h(L)=d(L)$ at the grounding line. 

Fig.~\ref{fig:timeseries_new} shows time series of 
randomly generated instances (random phases) of the incident wave field at peak periods $T_{\textnormal{p,in}}=20$\,s, $25$\,s and $30$\,s, along with corresponding time series of the ice shelf vibrations at the sample point ($x=2$\,km) and at the shelf front, where the shelf is clamped at the grounding line. 
The displacements are scaled by the significant wave height of the incident field, $H_{\textnormal{s,in}}$,
where
\begin{equation}
    H_{\textnormal{s,in}} = 4 \sqrt{
    \,\int_{0}^{\infty}S_{\textnormal{in}}(\omega)\wrt\omega.}
\end{equation}
The scaled shelf responses increase with increasing incident peak period, with the root mean square (RMS) of the shelf-vibration series relative to the incident-series RMS increasing by a factor of $1.7$--$2.5$ between $T_{\textnormal{p,in}}=20$\,s and $30$\,s.
The RMS of the series at the shelf front is greater than at the sample point ($x=2$\,km) by a factor of $5$--$9$.

An ocean-to-shelf transfer function for displacement is defined as
\begin{equation}
    \mathcal{T}_{\textnormal{d}}(x)=\frac{H_{\textnormal{s,sh}}(x)}{H_{\textnormal{s,in}}}
    \mathwhere
    H_{\textnormal{s,sh}}(x) =
    4\,\sqrt{\,\int_{0}^{\infty}
    S_{\textnormal{in}}(\omega)\,
    \left|\eta(x)\right|^{2}
    \wrt\omega},
\end{equation}
is the significant displacement height in the shelf at location $x$.
The transfer function at the sample point ($x=2$\,km) increases monotonically with increasing incident peak period for both clamped and hinged grounding-line conditions (Fig.~\ref{fig:transfer_new}a). 
The transfer function for the hinged shelf is greater than the clamped shelf by factor of 4.4 for the shortest incident peak period considered ($T_{\textnormal{p,in}}=15$\,s).
But the transfer function for the hinged shelf grows slower with increasing peak period than the clamped shelf, such that the transfer function for the clamped shelf slightly exceeds the hinged shelf at the longest period considered ($T_{\textnormal{p,in}}=30$\,s).

In analogy with the displacement transfer function, a transfer function for strain is defined as
\begin{equation}
    \mathcal{T}_{\textnormal{s}}(x)=
    \frac{H_{\textnormal{s},\textnormal{sh}}(x)}{H_{\textnormal{s,in}}}
    \mathwhere
    H_{\textnormal{s},\textnormal{sh}}(x) =
    4\,\sqrt{\,\int_{0}^{\infty}
    S_{\textnormal{in}}(\omega)\,
    \left|\epsilon(x)\right|^{2}
    \wrt\omega}
\end{equation}
is a significant strain height \cite{williams2013wavea,williams2013waveb}.
Fig.~\ref{fig:transfer_new}b shows the strain transfer function for the clamped and hinged shelves, evaluated at the sample point ($x=2$\,km).
Similarly to the displacement transfer function,
the strain transfer function is increases monotonically with increasing incident peak period. 
The strain transfer functions for the clamped and hinged shelves are similar, although the transfer function for the clamped shelf greater than the hinged shelf across the range of incident peak periods by a factor of 1.2--1.8, and a mean factor of 1.4.
}

\section{Conclusions}

A solution method has been developed for a model of ice shelf vibrations in response to ocean waves,
in which both the ice shelf thickness and the seabed beneath the shelf vary over distance, and the ice shelf and sub-shelf cavity are connected to the open ocean.
The method is based on a decomposition of the ice shelf vibrations into
free modes of uniform shelves, so that some of the most 
costly computations can be reused for different ice shelf geometries.

The method and model were used to study the influence of 
shoaling beneath the shelf,
\new{and} ice shelf thickening
\del{and viscoelastic damping}
on ice shelf vibrations in response to regular incident waves,
over a range of wave periods in the swell regime.
\new{Shoaling was found to increase flexural wavelengths and decrease amplitudes with distance along the shelf, 
    leading to resonant peaks in the maximum shelf displacement being shifted to longer periods and attaining slightly lower peak values.
    Similarly, shelf thickening was shown to increase wavelengths and decrease amplitudes along the shelf. 
    The effect of thickening on the resonant displacement peaks was found to be more complicated,
    but with a general trend for a severely thickening shelf to produce larger peaks and at longer periods than a uniform shelf with the same mean thickness.
    Shelf thickening was also shown to have a complicated effect on resonant strain peaks, with a mildly thickening shelf experiencing the largest strains.}
\begin{cross}
\begin{itemize}
    \item Shoaling was found to increase flexural wavelengths and decrease amplitudes with distance along in the ice shelf, 
    leading to resonant peaks in the maximum ice shelf displacement being shifted to longer periods and attaining slightly lower peak values.
    \item Similarly, ice shelf thickening was shown to increase wavelengths and decrease amplitudes along the shelf. 
    The effect on the resonant displacement peaks was found to be more complicated,
    but with a general trend for a severely thickening shelf to produce larger peaks and at longer periods than a uniform shelf with the same mean thickness.
    Shelf thickening was also shown to have a complicated effect on resonant strain peaks, with a mildly thickening shelf experiencing the largest strains.
    \item Viscoelastic damping (using the Kelvin--Voigt model) effectively eliminates resonances at short periods for a relatively small value of the damping parameter,
    but cannot eliminate long period resonances, even for a relatively large value of the damping parameter.
\end{itemize}
 \end{cross}
 \new{The responses of shelves with clamped and hinged grounding-line conditions were found to be similar, with small shifts in the hinged-shelf resonances to longer periods being the most notable difference.}

Results were generated for \new{long, thickening shelves over shoaling seabeds, forced by} irregular incident wave fields, 
using measurements at isolated points, 
which mimics recent experimental measurements\del{,
and an ocean-to-shelf transfer functions were defined in terms of significant displacement and strain heights.}\new{. Ocean-to-shelf transfer functions were defined in terms of significant displacement and strain heights.}
The key implications of the results are:
\begin{itemize}
    \begin{cross}
    \item 
    It is important to consider the ice shelf thickness profile, as using an average thickness leads to underestimates of displacements and strains at measurement points close to the shelf front, particularly for longer incident periods.
    \end{cross}
    \item
    The ice shelf response measured \del{at 10\,km} \new{$2$\,km} away from the shelf front may substantially underestimate the response closer to the shelf front, by up to an order of magnitude 
    in terms of the significant displacement height.
    \new{
    \item
    The relative displacement and strain (\ie transfer functions) close to the shelf front increase with increasing incident peak period.
    \item
    Shelves with clamped and hinged grounding-line conditions experience similar flexural strains close to the shelf front in response to irregular incident fields. However, this finding should be tested for a broader range of shelf/seabed geometries. 
    }
\end{itemize}
Moreover,
the results add to the growing body of evidence 
that ice shelves experience considerable vibrations in response to ocean swell.


\begin{thebibliography}{33}
\providecommand{\natexlab}[1]{#1}
\providecommand{\url}[1]{\texttt{#1}}
\expandafter\ifx\csname urlstyle\endcsname\relax
  \providecommand{\doi}[1]{doi: #1}\else
  \providecommand{\doi}{doi: \begingroup \urlstyle{rm}\Url}\fi

\bibitem[MacAyeal et~al.(2006)MacAyeal, Okal, Aster, Bassis, Brunt, Cathles,
  Drucker, Fricker, Kim, Martin, Okal, Sergienko, Sponsler, and
  Thom]{macayeal2006transoceanic}
D.~R. MacAyeal, E.~A. Okal, R.~C. Aster, J.~N. Bassis, K.~M. Brunt, L.~M.
  Cathles, R.~Drucker, H.~A. Fricker, Y.-J. Kim, S.~Martin, M.~H. Okal, O.~V.
  Sergienko, M.~P. Sponsler, and J.~E. Thom.
\newblock Transoceanic wave propagation links iceberg calving margins of
  {A}ntarctica with storms in tropics and {N}orthern {H}emisphere.
\newblock \emph{Geophys. Res. Lett.}, 33\penalty0 (17), 2006.

\bibitem[Cathles et~al.(2009)Cathles, Okal, and MacAyeal]{cathles2009seismic}
L.~M. Cathles, E.~A. Okal, and D.~R. MacAyeal.
\newblock Seismic observations of sea swell on the floating {R}oss {I}ce
  {S}helf, {A}ntarctica.
\newblock \emph{J. Geophys. Res.: Earth Surface}, 114\penalty0 (F2), 2009.

\bibitem[Bromirski et~al.(2010)Bromirski, Sergienko, and
  MacAyeal]{bromirski2010transoceanic}
P.~D. Bromirski, O.~V. Sergienko, and D.~R. MacAyeal.
\newblock Transoceanic infragravity waves impacting {A}ntarctic ice shelves.
\newblock \emph{Geophys. Res. Lett.}, 37\penalty0 (2), 2010.

\bibitem[Bromirski and Stephen(2012)]{Bro&Ste12}
P.~D. Bromirski and R.~A. Stephen.
\newblock {Response of the {R}oss {I}ce {S}helf, {A}ntarctica, to ocean
  gravity-wave forcing}.
\newblock \emph{Annals Glaciol.}, 53\penalty0 (60):\penalty0 163--172, 2012.

\bibitem[Bromirski et~al.(2015)Bromirski, Diez, Gerstoft, Stephen, Bolmer,
  Wiens, Aster, and Nyblade]{Broetal15}
P.~D. Bromirski, A.~Diez, P.~Gerstoft, R.~A. Stephen, T.~Bolmer, D.~A. Wiens,
  R.~Aster, and N.~Nyblade.
\newblock {Ross {I}ce {S}helf vibrations}.
\newblock \emph{Geophys. Res. Lett.}, 42:\penalty0 7589--7597, 2015.

\bibitem[Bromirski et~al.(2017)Bromirski, Chen, Stephen, Gerstoft, Arcas, Diez,
  Aster, Wiens, and Nyblade]{bromirski2017tsunami}
P.~D. Bromirski, Z.~Chen, R.~A. Stephen, P.~Gerstoft, D.~Arcas, A.~Diez, R.~C.
  Aster, D.~A. Wiens, and A.~Nyblade.
\newblock Tsunami and infragravity waves impacting antarctic ice shelves.
\newblock \emph{J. Geophys. Res.: Oceans}, 2017.

\bibitem[Chen et~al.(2019)Chen, Bromirski, Gerstoft, Stephen, Lee, Yun,
  Olinger, Aster, Wiens, and Nyblade]{chen2019ross}
Z.~Chen, P.~D. Bromirski, P.~Gerstoft, R.~A. Stephen, W.~S. Lee, S.~Yun, S.~D.
  Olinger, R.~C. Aster, D.~A. Wiens, and A.~A. Nyblade.
\newblock {R}oss {I}ce {S}helf icequakes associated with ocean gravity wave
  activity.
\newblock \emph{Geophys. Res. Lett.}, 46\penalty0 (15):\penalty0 8893--8902,
  2019.

\bibitem[Brunt et~al.(2011)Brunt, Okal, and Mac{A}yeal]{brunt2011antarctic}
K.~M. Brunt, E.~A. Okal, and D.~R. Mac{A}yeal.
\newblock Antarctic ice-shelf calving triggered by the {H}onshu ({J}apan)
  earthquake and tsunami, {M}arch 2011.
\newblock \emph{J. Glaciol.}, 57\penalty0 (205):\penalty0 785--788, 2011.

\bibitem[Massom et~al.(2018)Massom, Scambos, Bennetts, Reid, Squire, and
  Stammerjohn]{massom_etal18}
R.~A. Massom, T.~A. Scambos, L.~G. Bennetts, P.~Reid, V.~A. Squire, and S.~E.
  Stammerjohn.
\newblock Antarctic ice shelf disintegration triggered by sea ice loss and
  ocean swell.
\newblock \emph{Nature}, 558:\penalty0 383--389, 2018.

\bibitem[Gudmundsson(2013)]{gudmundsson2013}
G.~H. Gudmundsson.
\newblock Ice-shelf buttressing and the stability of ice sheets.
\newblock \emph{Cryosphere}, 7:\penalty0 647--655, 2013.

\bibitem[Scambos et~al.(2004)Scambos, Bohlander, Shuman, and
  Skvarca]{scambos2004glacier}
T.~A. Scambos, J.~A. Bohlander, C.~A. Shuman, and P.~Skvarca.
\newblock Glacier acceleration and thinning after ice shelf collapse in the
  larsen b embayment, antarctica.
\newblock \emph{Geophys. Res. Lett.}, 31\penalty0 (18), 2004.

\bibitem[Holdsworth and Glynn(1978)]{holdsworth1978iceberg}
G.~Holdsworth and J.~E. Glynn.
\newblock Iceberg calving from floating glaciers by a vibrating mechanism.
\newblock \emph{Nature}, 274:\penalty0 464--466, 1978.

\bibitem[Holdsworth and Glynn(1981)]{holdsworth1981mechanism}
G.~Holdsworth and J.~E. Glynn.
\newblock A mechanism for the formation of large icebergs.
\newblock \emph{J. Geophys. Res.: Oceans}, 86\penalty0 (C4):\penalty0
  3210--3222, 1981.

\bibitem[Vinogradov and Holdsworth(1985)]{vinogradov1985oscillation}
O.~G. Vinogradov and G.~Holdsworth.
\newblock Oscillation of a floating glacier tongue.
\newblock \emph{Cold Reg. Sci. Tech.}, 10\penalty0 (3):\penalty0 263--271,
  1985.

\bibitem[Fox and Squire(1991)]{fox1991coupling}
C.~Fox and V.~A. Squire.
\newblock Coupling between the ocean and an ice shelf.
\newblock \emph{Annals Glaciol.}, 15:\penalty0 101--108, 1991.

\bibitem[Kalyanaraman et~al.(2020)Kalyanaraman, Meylan, Bennetts, and
  Lamichhane]{kalyanaraman_coupled_2020}
B.~Kalyanaraman, M.~H. Meylan, L.~G. Bennetts, and B.~P. Lamichhane.
\newblock A coupled fluid--elasticity model for the wave forcing of an
  ice-shelf.
\newblock \emph{J. Fluids Struct.}, 97:\penalty0 103074, 2020.

\bibitem[Holdsworth(1977)]{holdsworth1977tidal}
G.~Holdsworth.
\newblock Tidal interactions with ice shelves.
\newblock \emph{Ann. Geophys.}, 33\penalty0 (1--2):\penalty0 133--146, 1977.

\bibitem[Sayag and Worster(2013)]{sayag_elastic_2013}
R.~Sayag and M.~G. Worster.
\newblock Elastic dynamics and tidal migration of grounding lines modify
  subglacial lubrication and melting.
\newblock \emph{Geophys. Res. Lett.}, 40\penalty0 (22):\penalty0 5877--5881,
  2013.

\bibitem[Rosier et~al.(2017)Rosier, Marsh, Rack, Gudmundsson, Wild, and
  Ryan]{rosier_interpretation_2017}
S.~H.~R. Rosier, O.~J. Marsh, W.~Rack, G.~H. Gudmundsson, C.~T. Wild, and
  M.~Ryan.
\newblock On the interpretation of ice-shelf flexure measurements.
\newblock \emph{J. Glaciol.}, 63\penalty0 (241):\penalty0 783--791, 2017.

\bibitem[Sergienko(2013)]{Ser13}
O.~V. Sergienko.
\newblock {Normal modes of a coupled ice-shelf/sub-ice-shelf cavity system}.
\newblock \emph{J. Glaciol.}, 59\penalty0 (213):\penalty0 76--80, 2013.
\newblock \doi{10.3189/2013JoG12J096}.

\bibitem[Meylan et~al.(2017)Meylan, Bennetts, Hosking, and
  Catt]{meylan_etal17_annals}
M.~H. Meylan, L.~G. Bennetts, R.~J. Hosking, and E.~Catt.
\newblock On the calculation of normal modes of a coupled
  ice-shelf/sub-ice-shelf cavity system.
\newblock \emph{J. Glaciol.}, 63\penalty0 (240):\penalty0 751--754, 2017.

\bibitem[Sergienko(2010)]{Sergienko2010}
O.~V. Sergienko.
\newblock {Elastic response of floating glacier ice to impact of long-period
  ocean waves}.
\newblock \emph{J. Geophys. Res.: Earth Surface}, 115\penalty0 (4):\penalty0
  1--16, 2010.
\newblock \doi{10.1029/2010JF001721}.

\bibitem[Sergienko(2017)]{sergienko2017behavior}
O.~V. Sergienko.
\newblock Behavior of flexural gravity waves on ice shelves: Application to the
  {R}oss {I}ce {S}helf.
\newblock \emph{J. Geophys. Res.: Oceans}, 2017.

\bibitem[Papathanasiou et~al.(2015{\natexlab{a}})Papathanasiou, Karperaki,
  Theotokoglou, and Belibassakis]{papathanasiou_higher_2015}
T.~K. Papathanasiou, A.~Karperaki, E.~E. Theotokoglou, and K.~A. Belibassakis.
\newblock A higher order {FEM} for time-domain hydroelastic analysis of large
  floating bodies in an inhomogeneous shallow water environment.
\newblock \emph{Proc. Roy. Soc. A}, 471\penalty0 (2173):\penalty0 20140643,
  2015{\natexlab{a}}.

\bibitem[Papathanasiou et~al.(2015{\natexlab{b}})Papathanasiou, Karperaki,
  Theotokoglou, and Belibassakis]{papathanasiou2015hydroelastic}
T.~K. Papathanasiou, A.~E. Karperaki, E.~E. Theotokoglou, and K.~A.
  Belibassakis.
\newblock Hydroelastic analysis of ice shelves under long wave excitation.
\newblock \emph{Nat. Hazard Earth Sys.}, 15\penalty0 (8):\penalty0 1851--1857,
  2015{\natexlab{b}}.

\bibitem[Papathanasiou et~al.(2019)Papathanasiou, Karperaki, and
  Belibassakis]{papathanasiou2019resonant}
T.~K. Papathanasiou, A.~E. Karperaki, and K.~A. Belibassakis.
\newblock On the resonant hydroelastic behaviour of ice shelves.
\newblock \emph{Ocean Model.}, 133:\penalty0 11--26, 2019.

\bibitem[Kalyanaraman et~al.(2019)Kalyanaraman, Bennetts, Lamichhane, and
  Meylan]{kalyanaraman2019shallow}
B.~Kalyanaraman, L.~G. Bennetts, B.~Lamichhane, and M.~H. Meylan.
\newblock On the shallow-water limit for modelling ocean-wave induced ice-shelf
  vibrations.
\newblock \emph{Wave Motion}, 90:\penalty0 1--16, 2019.

\bibitem[Ilyas et~al.(2018)Ilyas, Meylan, Lamichhane, and
  Bennetts]{ilyas2018time}
M.~Ilyas, M.~H. Meylan, B.~Lamichhane, and L.~G. Bennetts.
\newblock Time-domain and modal response of ice shelves to wave forcing using
  the finite element method.
\newblock \emph{J. Fluids Struct.}, 80:\penalty0 113--131, 2018.

\bibitem[Fretwell et~al.(2013)Fretwell, Pritchard, Vaughan, Bamber, Barrand,
  Bell, Bianchi, Bingham, Blankenship, Casassa, Catania, Callens, Conway, Cook,
  Corr, Damaske, Damm, Ferraccioli, Forsberg, Fujita, Gim, Gogineni, Griggs,
  Hindmarsh, Holmlund, Holt, Jacobel, Jenkins, Jokat, Jordan, King, Kohler,
  Krabill, Riger-Kusk, Langley, Leitchenkov, Leuschen, Luyendyk, Matsuoka,
  Mouginot, Nitsche, Nogi, Nost, Popov, Rignot, Rippin, Rivera, Roberts, Ross,
  Siegert, Smith, Steinhage, Studinger, Sun, Tinto, Welch, Wilson, Young,
  Xiangbin, and Zirizzotti]{bedmap2}
P.~Fretwell, H.~D. Pritchard, D.~G. Vaughan, J.~L. Bamber, N.~E. Barrand,
  R.~Bell, C.~Bianchi, R.~G. Bingham, D.~D. Blankenship, G.~Casassa,
  G.~Catania, D.~Callens, H.~Conway, A.~J. Cook, H.~F.~J. Corr, D.~Damaske,
  V.~Damm, F.~Ferraccioli, R.~Forsberg, S.~Fujita, Y.~Gim, P.~Gogineni, J.~A.
  Griggs, R.~C.~A. Hindmarsh, P.~Holmlund, J.~W. Holt, R.~W. Jacobel,
  A.~Jenkins, W.~Jokat, T.~Jordan, E.~C. King, J.~Kohler, W.~Krabill,
  M.~Riger-Kusk, K.~A. Langley, G.~Leitchenkov, C.~Leuschen, B.~P. Luyendyk,
  K.~Matsuoka, J.~Mouginot, F.~O. Nitsche, Y.~Nogi, O.~A. Nost, S.~V. Popov,
  E.~Rignot, D.~M. Rippin, A.~Rivera, J.~Roberts, N.~Ross, M.~J. Siegert, A.~M.
  Smith, D.~Steinhage, M.~Studinger, B.~Sun, B.~K. Tinto, B.~C. Welch,
  D.~Wilson, D.~A. Young, C.~Xiangbin, and A.~Zirizzotti.
\newblock Bedmap2: improved ice bed, surface and thickness datasets for
  antarctica.
\newblock \emph{Cryosphere}, 7\penalty0 (1):\penalty0 375--393, 2013.

\bibitem[Vaughan(1995)]{vaughan_tidal_1995}
D.~G. Vaughan.
\newblock Tidal flexure at ice shelf margins.
\newblock \emph{J. Geophys. Res.: Solid Earth}, 100\penalty0 (B4):\penalty0
  6213--6224, 1995.

\bibitem[Kahma et~al.(2005)Kahma, Hauser, Krogstad, Lehner, Monbaliu, and
  Wyatt]{Kahma2005}
K~Kahma, D~Hauser, H.~E. Krogstad, S.~Lehner, J.~A.~J. Monbaliu, and L.~R.
  Wyatt.
\newblock \emph{{Measuring and analysing the directional spectra of ocean
  waves}}.
\newblock Luxembourg: Office for Official Publications of the European
  Communities, 2005.
\newblock ISBN 9289800038.

\bibitem[Williams et~al.(2013{\natexlab{a}})Williams, Bennetts, Squire, Dumont,
  and Bertino]{williams2013wavea}
T.~D. Williams, L.~G. Bennetts, V.~A. Squire, D.~Dumont, and L.~Bertino.
\newblock Wave--ice interactions in the marginal ice zone. {P}art 1:
  {T}heoretical foundations.
\newblock \emph{Ocean Model.}, 71:\penalty0 81--91, 2013{\natexlab{a}}.

\bibitem[Williams et~al.(2013{\natexlab{b}})Williams, Bennetts, Squire, Dumont,
  and Bertino]{williams2013waveb}
T.~D. Williams, L.~G. Bennetts, V.~A. Squire, D.~Dumont, and L.~Bertino.
\newblock Wave--ice interactions in the marginal ice zone. {P}art 2:
  {N}umerical implementation and sensitivity studies along {1D} transects of
  the ocean surface.
\newblock \emph{Ocean Model.}, 71:\penalty0 92--101, 2013{\natexlab{b}}.

\end{thebibliography}
\end{document}